\documentclass[lettersize,journal]{IEEEtran}
\usepackage{cite}
\usepackage[dvipsnames, table]{xcolor}
\usepackage{colortbl}

\makeatletter
\def\@IEEEstripcolor#1{\color{#1}}
\makeatother

\usepackage{amsmath,amssymb,amsfonts}
\usepackage{bm}
\usepackage{optidef}
\usepackage{mathptmx} 
\DeclareMathOperator*{\argmax}{arg\,max}
\DeclareMathOperator*{\argmin}{arg\,min}

\usepackage{algorithm}
\usepackage{algpseudocode}

\usepackage{booktabs}
\usepackage{makecell}
\usepackage{multirow}
\usepackage{tabularx}
\usepackage{array}
\usepackage{hhline}
\usepackage{arydshln}

\usepackage{adjustbox}

\usepackage{graphicx}
\usepackage{wrapfig}
\usepackage{adjustbox}
\usepackage{caption}
\usepackage{subcaption}

\usepackage{textcomp}
\usepackage{xurl}
\usepackage{indentfirst}
\usepackage{lipsum}

\usepackage{soul}        
\usepackage{soulpos}     
\usepackage[normalem]{ulem} 

\usepackage{amssymb}
\usepackage{pifont}

\newcommand{\cmark}{\ding{51}}
\newcommand{\xmark}{\ding{55}}

\newif\ifshownormalchanges
\shownormalchangesfalse 

\newif\ifshowsuperchanges
\showsuperchangesfalse   

\newcommand{\added}[1]{%
  \ifshownormalchanges
    {\textcolor{blue}{#1}}%
  \else
    #1%
  \fi
}

\newcommand{\deleted}[1]{%
  \ifshownormalchanges
    {\textcolor{red}{\sout{#1}}}%
  \fi
}

\newcommand{\superadded}[1]{%
  \ifshowsuperchanges
    {\textcolor{blue}{#1}}%
  \else
    #1%
  \fi
}

\usepackage[hidelinks]{hyperref}

\begin{document}

\title{RepTC: Representation-Aware Optimization for Efficient Traffic Classification on Edge IoT Devices}
\author{%
Adel Chehade,~\IEEEmembership{Graduate Student Member,~IEEE,}
Edoardo Ragusa,~\IEEEmembership{Member,~IEEE,}
Paolo Gastaldo,~\IEEEmembership{Member,~IEEE,}
and Rodolfo Zunino
\thanks{The authors are with the Department of Naval, Electrical, Electronic and Telecommunications Engineering (DITEN), University of Genoa, 16126 Genoa, Italy (e-mail: adel.chehade@edu.unige.it; edoardo.ragusa@unige.it; paolo.gastaldo@unige.it; rodolfo.zunino@unige.it).}

\thanks{This work was partially supported by project SERICS (PE00000014) under the MUR National Recovery and Resilience Plan funded by the European Union - NextGenerationEU.}
}


\maketitle

\begin{abstract}
Traffic classification (TC) is crucial to secure Internet of Things (IoT) networks, whose edge nodes often operate under privacy, bandwidth, and energy constraints.
Yet, encrypted payloads and limited computing power make accurate, real-time TC a challenging task.
Existing learning-based TC approaches often fix the input configuration a priori, even though it directly influences both predictive performance and computational cost.
This paper presents RepTC, a representation-aware hardware-constrained strategy that addresses session-level TC 
through joint model and input optimization. The method co-optimizes network architecture, session length, and header preprocessing; this enables joint control of model complexity, input scale, and data representation within a unified resource-constrained design space.
The proposed approach enforces microcontroller-class constraints on memory, model size, and computation, and yields compact models deployable on low-power edge devices.
A gateway monitors traffic by aggregating sessions and either performs inference locally or offloads to a low-power edge node. \deleted{The research presented in the paper validates both scenarios through experiments on a Raspberry Pi 3B\texttt{+} and STM32 Nucleo-F401RE, demonstrating limited inference latencies (0.63--9.1\,ms) and minimal energy requirements per session (below 0.83\,mJ).}\superadded{Both scenarios are validated on heterogeneous embedded hardware, including a Raspberry Pi 3B\texttt{+}, STM32 Nucleo-F401RE, and XIAO
ESP32-C3, spanning Cortex-A, Cortex-M, and RISC-V processor architectures; measured inference latency ranges from 0.63 to 18.59\,ms, with MCU-side
inference energy between 0.50 and 1.41\,mJ per session.}
RepTC yields configurations that achieve high accuracy on a variety of established benchmarks: 96.21\% on ISCX VPN-nonVPN, 99.62\% on USTC-TFC2016, and 99.97\% on Edge-IIoTset; at the same time, model size and computational requirements were reduced by up to three orders of magnitude compared with state-of-the-art methods.
The results show that representation-aware optimization can improve efficiency while preserving competitive classification performance.

\end{abstract}



\maketitle




\section{Introduction}

\IEEEPARstart{T}{he} integration of Internet of Things (IoT) devices in mission-critical applications~\cite{xu2024cascaded, yao2019capsule} has increased the relevance of security in data  exchange. Traffic classification (TC) has become a crucial requirement to secure networks, detect anomalies, and optimize resources~\cite{nas_wang2024toward_predictor,gallo2021fenxi}.
Encrypted communication obfuscates payload and weakens the effectiveness of conventional TC approaches based on payload inspection and static port analysis~\cite{survey_DONG2024128444}.
Classifiers must therefore rely on structural and behavioral patterns reflected in network traffic. Accordingly, proper processing of the input data can amplify the information contained in the data stream, thereby simplifying the classifier’s task. Despite this, while tremendous efforts have been devoted to the design of classifiers\cite{our_chehade2025tnsm,nas_zhang2023automatic,nas_etc_malekghaini2023automl4etc}, the automatic optimization of input representations remains relatively underexplored.
The approach presented in this paper adopts a representation-aware optimization strategy that jointly optimizes deep neural networks (DNNs) and their input representation, as these components ultimately determine a system's footprint, computational cost, and classification performance in edge deployments. This formulation targets compact neural models operating under strict edge constraints. The considered platform is a lightweight  node (e.g., an embedded microcontroller) that operates alongside sensing devices in the network.

On-device classification avoids transmission to external servers and is thus crucial to reduce latency, reduce bandwidth usage, and ensure security~\cite{our_chehade2024tiny, gallo2021fenxi}. In IoT applications, real-time response and data privacy directly affect system reliability and effectiveness.  
Edge deployment poses some inherent challenges: limited compute requires lightweight models, encrypted traffic exhibits complex patterns, and classification errors can be costly (e.g., due to security breaches or resource misallocation)~\cite{survey_DONG2024128444,gallo2021fenxi}. Deep learning (DL) provides a viable solution, since DL models can learn discriminative patterns directly from raw traffic, thus avoiding pre-computed feature engineering~\cite{nas_etc_malekghaini2023automl4etc,seydali2023cbs}. Conventional DL architectures, however, far exceed the hardware limits of IoT devices. 
This challenge is not only architectural; for compact models, performance and efficiency depend critically on the input configuration, which influences both the information available to the network and the hardware cost of inference ~\cite{lin2020mcunet, njor2025fast}.

In the context of TC, this limitation becomes more evident, as optimizing the architecture alone may not be sufficient. Due to the strong structural dependencies and privacy constraints of traffic data~\cite{lotfollahi2020deep,yao2019identification}, the choice of input scale and representation plays a central role in determining both predictive performance and computational cost. Considering these factors separately limits the adaptability of existing solutions.

Motivated by that interdependence, we incorporate input representation optimization into the design process, resulting in a unified approach that jointly optimizes all components relevant to the design of hardware-efficient models for TC.

The proposed approach (i) defines a design space of compact DNN architectures along with input representation parameters, (ii) explores this space to identify high-performing configurations, and (iii) leverages a hardware-aware cost function to balance predictive accuracy and resource usage; we further analyze the resulting search space to provide practical insight into the joint design trade-offs.

Specifically, TC applies at different granularity levels: packets offer fine-grained views but lack context; flows aggregate unidirectional traffic, whereas sessions capture bidirectional exchanges and enable deeper structural understanding \cite{li2024l2}. 
For each level, DL models typically rely on a fixed input size, which determines how much traffic is processed per sample. In many session- and flow-based studies, 784-byte inputs are adopted \cite{wang2017end, wang2017malware,lu2021iclstm,zou2024novel_residual,only_header_cui2022only}, while packet-level methods often use 1500-byte samples \cite{lotfollahi2020deep,seydali2023cbs}. These settings, however, are usually chosen a priori and often lack some principled analysis; whether shorter inputs might improve efficiency or predictive performance remains underexplored.

Traffic preprocessing also varies in how header fields are handled (e.g., IP, MAC, ports) \cite{our_chehade2025tnsm,cui2019session}. Transformations (such as anonymization, zeroing, or removal) aim to reduce overfitting and to address privacy concerns, especially when deploying models in sensitive settings. Yet, their downstream impact on classification accuracy is rarely quantified. To the best of our knowledge, preprocessing and input representation have never been co-optimized with architectural design.

The proposed RepTC approach addresses these limitations by formulating a joint hardware-aware optimization strategy over model structure, input length, and header preprocessing.
This joint formulation enables flexible trade-offs among accuracy, efficiency, and deployment cost.
In line with many state-of-the-art methods, RepTC focuses on session-level classification and exploits bidirectional exchanges to capture the structural semantics of network traffic.

The experimental evaluation of this approach involved three datasets: ISCX VPN-nonVPN~\cite{iscxvpnnonvpn} (encrypted and non-encrypted traffic); USTC-TFC2016~\cite{wang2017malware} (network applications and malware), and Edge-IIoTset~\cite{edgeiiot_ferrag2022edge} (industrial traffic with intrusion scenarios). The results show that adapting input length and header preprocessing together with the model architecture improves classification accuracy while reducing resource requirements; further analysis indicates that satisfactory configurations under strict hardware constraints emerge only when these components are optimized jointly.
This positions input representation as a key factor in the hardware-constrained design of TC models for edge deployment.

The main contributions of the presented research can be summarized as follows:

\begin{itemize}
    \item The RepTC approach formulates TC model design as a joint optimization problem over DNN architecture, input size, and header preprocessing. This unified design space yields lightweight, high-performing configurations for edge devices; it bypasses the need for manual tuning and outperforms static-input methods.

    \item When considering state-of-the-art TC on encrypted traffic and industrial IoT benchmarks, the new models generated using the proposed approach turn out to be up to hundreds of times smaller in parameters and tensor footprint, and up to thousands of times more efficient in floating-point operations (FLOPs).

    \item An in-depth characterization of the RepTC search space quantifies how architectural and input factors relate to accuracy through mutual information (MI), correlation, and analysis of variance (ANOVA); the findings reveal significant interactions among network depth, receptive fields, and input configuration, supporting their joint consideration under hardware constraints.

    \item In conjunction with those reductions in computational and memory requirements, the RepTC strategy could match or outperform state-of-the-art results, scoring 96.21\% on ISCX VPN-nonVPN, 99.62\% on USTC-TFC2016, and 99.97\% on Edge-IIoTset.

   \deleted{\textbullet\ The method underwent full deployment on Raspberry Pi and
STM32 hardware platforms, confirming the reliable support for end-to-end
session-level TC on IoT devices.}

\deleted{\textbullet\ The energy consumption for quantized inference on STM32
was under 1\,mJ per session, thus proving that the method could fit
battery-operated, ultra-low-power nodes.}

\item \superadded{Extensive real-hardware validation demonstrates end-to-end
session-level TC across heterogeneous embedded platforms, including Raspberry
Pi, STM32 Cortex-M4, and ESP32 RISC-V targets; inference latency ranges from
0.63 to 18.59\,ms, with MCU-side inference energy of
0.50--1.41\,mJ per session.}

\item \superadded{The proposed RepTC framework is publicly available at
\url{https://github.com/SEAlab-unige/RepTC}.}

\end{itemize}

\section{Related Work}

\subsection{Conventional Traffic Classification Methods}

Conventional TC mainly relies on port-based methods and deep packet inspection (DPI). Port-based approaches map traffic to applications through port/service numbers (e.g., HTTPS on port 443), but their reliability has somewhat declined with dynamic ports, tunneling, and obfuscation~\cite{ports_niu2019heuristic,survey_DONG2024128444}. DPI inspects packet payloads through signature matching, achieving high accuracy on plain traffic, but it raises privacy concerns, introduces notable overhead, and becomes considerably less effective on encrypted traffic~\cite{pacheco2018towards,survey_DONG2024128444}.

Machine learning (ML) approaches based on handcrafted features and classical classifiers, such as Random Forests and k-nearest neighbors (k-NN), have shown effectiveness in tasks including intrusion detection and application identification~\cite{ml_rf_supervised_zhai2018random,iscxvpnnonvpn,App_yamansavascilar2017application,edgeiiot_ferrag2022edge}. However, their reliance on manual feature engineering limits scalability and adaptability. Dimensionality reduction techniques such as principal component analysis (PCA) and t-distributed stochastic neighbor embedding (t-SNE)~\cite{pca_gudla2024novel} have also been explored, but often introduce bias and remain sensitive to traffic variability.

\subsection{Deep Learning for Traffic Classification}

The success of Deep Learning for TC mostly stems from its ability to learn hierarchical features directly from raw traffic. 
In the seminal work~\cite{wang2017end}, one-dimensional convolutional neural networks (1D-CNNs) outperform two-dimensional CNNs (2D-CNNs) at session-level encrypted traffic classification (ETC). A related study in~\cite{wang2017malware} applies the same 1D-CNN structure for session-based malware detection, achieving remarkable results. Several subsequent works thus adopted the same 784-byte, fixed-length representation. For instance, the study in~\cite{wang2020encrypted} combines CNNs with stacked autoencoders; others explore architectures like Inception–long short-term memory (Inception-LSTM) networks~\cite{lu2021iclstm} or incorporate attention and squeeze-and-excitation modules~\cite{zou2024novel_residual}.
Those models mainly prioritize accuracy and are often resource-demanding, with limited consideration of efficiency or adaptability.

Other approaches operate at the flow or packet level. Flow-based methods vary in design: some handle raw bytes as time series~\cite{yao2019identification}, others use 784-byte header-derived inputs~\cite{only_header_cui2022only} or extract statistical features~\cite{selem2025deep,alshehri2024self,hassini2024end,ullah2024attention}. One should anyway consider that feature extraction can delay inference and reduce adaptability.
Packet-level methods often involve 1500-byte inputs, such as the model in~\cite{lotfollahi2020deep} for ETC. In malware detection, the study in~\cite{malicious_packets_zhou2023identification} includes a 1D-CNN with attention, while the authors in~\cite{ieeeaccess_wang2020deep} integrate CNNs and gated recurrent units (GRUs). 
Since each packet is processed individually, these methods tend to perform frequent classifications at inference time, thereby increasing computational overhead.

Hybrid models operate at multiple granularity levels. The approach~\cite{cui2019session} segments sessions into packet groups, whereas~\cite{seydali2023cbs} combines session-level and packet-level inputs. 
These methods prove quite demanding computationally and
depend on offline processing; this limits real-time deployment
and still lacks extensive research in that sense.

Several studies have also explored alternative representations of raw traffic, such as audio-based encodings or graph-based formulations of packet or byte relationships~\cite{chen2026trafficaudio,li2026bpf_gnn}. These methods can reveal additional structural patterns, but they also introduce extra preprocessing stages or higher computational overhead.

Overall, prior DL studies differ not only in architecture and granularity, but also in how traffic is represented and preprocessed. Yet these choices are usually fixed in advance, with limited analysis of how they affect accuracy, efficiency, or downstream deployment constraints.

\subsection{Architecture Search and Representation Design}
Recent work has explored automated methods to improve the efficiency of DL models for TC.
One direction explores neural architecture search (NAS), which automates neural network design for a given task and dataset~\cite{1000papersNAS}.  
NAS has been applied extensively in domains such as computer vision and language modeling~\cite{chitty2023neural_ieee-access}, but its use in network TC remains relatively limited. Existing applications include ETC using packet headers~\cite{nas_etc_malekghaini2023automl4etc}, malware detection via proximal iteration~\cite{nas_zhang2023automatic}, and proxy-based accuracy prediction for intrusion detection~\cite{nas_wang2024toward_predictor}. These approaches generally optimize the architecture under fixed input settings and without explicit hardware constraints.

Hardware-aware neural architecture search (HW-NAS) incorporates hardware metrics such as parameter count, memory usage, and FLOPs into the design process~\cite{1000papersNAS,ragusa2024compression,ragusa2024combining}. 
Embedded-ML research has further explored architecture--compiler co-optimization~\cite{chen2022framework} and joint optimization of input configuration and network architecture~\cite{njor2025fast}.
\added{Among approaches targeting MCU-class deployment, representative HW-NAS frameworks include MCUNet~\cite{lin2020mcunet} and $\mu$NAS~\cite{liberis2021munas}. MCUNet couples architecture search with a lightweight inference engine and adapts architectural dimensions and input resolution to device constraints, whereas $\mu$NAS constrains architecture search according to RAM, persistent-storage, and processor requirements.} 
However, such approaches have seen limited application to network TC, where representation design involves interdependent choices such as input length and protocol-dependent header preprocessing, \added{with direct consequences for the traffic information retained for classification.}

\added{Within TC,} HW-NAS has been applied to session-level classification~\cite{our_chehade2024tiny,our_chehade2025iscc}, with~\cite{our_chehade2025iscc} even demonstrating deployment on microcontrollers. These designs, however, still relied on fixed input configurations, 
leaving a significant portion of the accuracy–efficiency trade-off determined by manual representation choices.
The study in~\cite{our_chehade2025tnsm} further analyzed how input length and header transformations affect accuracy and efficiency, but considered these only in post-hoc sensitivity analysis rather than integrating them into the search itself.
This separation restricts the exploration of interactions between input representation and model design.


\added{Table~\ref{tab:general_nas_comparison} summarizes these approaches with respect to hardware awareness,
architecture search, input-scale search, input-configuration search, and
application to TC.}

\added{MCUNet~\cite{lin2020mcunet} adapts input scale together with the
architecture, while Data-aware NAS~\cite{njor2025fast} additionally searches
input-data configuration. These methods target non-traffic domains and do not
consider traffic-specific variables such as input length and
protocol-dependent header preprocessing. Prior TC-specific
HW-NAS~\cite{our_chehade2025tnsm}, in contrast, operates directly on traffic
data but keeps these representation choices fixed during architecture search.}
\superadded{MicroNAS~\cite{king2025micronas} similarly focuses on
hardware-constrained architecture search without extending the search to the
input configuration.}

\begin{table}[htbp]
\centering
\caption{\added{Representative hardware-aware search comparison.}}
\label{tab:general_nas_comparison}

\renewcommand{\arraystretch}{0.90}
\small
\setlength{\tabcolsep}{3.0pt}

\begin{adjustbox}{max width=\columnwidth}
\begin{tabular}{
>{\centering\arraybackslash}c|
>{\centering\arraybackslash}c||
>{\centering\arraybackslash}c|
>{\centering\arraybackslash}c|
>{\centering\arraybackslash}c||
>{\centering\arraybackslash}c
}
\hline

\textbf{Method}
& \makecell{\textbf{HW}\\\textbf{Aware}}
& \makecell{\textbf{Arch.}\\\textbf{Search}}
& \makecell{\textbf{Input-Scale}\\\textbf{Search}}
& \makecell{\textbf{Input Config.}\\\textbf{Search}}
& \makecell{\textbf{TC}\\\textbf{Specific}}
\\
\hline

MCUNet~\cite{lin2020mcunet}
& \cmark
& \cmark
& \cmark
& \xmark
& \xmark
\\
\hline

$\mu$NAS~\cite{liberis2021munas}
& \cmark
& \cmark
& \xmark
& \xmark
& \xmark
\\
\hline

MicroNAS~\cite{king2025micronas}
& \cmark
& \cmark
& \xmark
& \xmark
& \xmark
\\
\hline

Data-aware NAS~\cite{njor2025fast}
& \cmark
& \cmark
& \cmark
& \cmark
& \xmark
\\
\hline

TC HW-NAS~\cite{our_chehade2025tnsm}
& \cmark
& \cmark
& \xmark
& \xmark
& \cmark
\\
\hline

\textbf{RepTC}
& \textbf{\cmark}
& \textbf{\cmark}
& \textbf{\cmark}
& \textbf{\cmark}
& \textbf{\cmark}
\\
\hline

\end{tabular}
\end{adjustbox}
\end{table}

\added{Complementing this general comparison, }Table~\ref{tab:tc_representation_comparison} summarizes representative raw-traffic-based TC approaches with respect to traffic granularity (Gran.), architecture design, input representation (including preprocessing strategies and input shape choices), and hardware awareness.

The comparison shows that while architectures vary across studies, preprocessing rules and input sizes are typically fixed beforehand. Header-level operations such as anonymization (anonym.), Ethernet removal (Eth. removal), or field zeroing are usually chosen manually, and commonly used input sizes (e.g., 784-byte sessions or 1500-byte packets) are adopted from convention rather than optimized systematically.

Thus, while prior efforts improved architectural efficiency and explored
hardware- and data-aware optimization, the coupled impact of traffic-specific input representation on both accuracy and hardware cost, and its interaction with the resulting architecture, remains underexplored.
This work addresses that gap by integrating input length and header preprocessing directly into the optimization process, enabling joint adaptation of the model and its input under strict device constraints.

\begin{table}[t]
\centering
\caption{Comparison of representative raw-traffic-based TC approaches with respect to traffic granularity, architecture design, input representation, and hardware awareness.}
\label{tab:tc_representation_comparison}
\Huge
\renewcommand{\arraystretch}{1.08}
\resizebox{\columnwidth}{!}{%
\begin{tabular}{c |c| c| c| c| c| c| c| c}
\hline
\multicolumn{1}{c|}{} & \multicolumn{1}{c|}{} & \multicolumn{2}{c|}{\textbf{Architecture Design}} & \multicolumn{4}{c|}{\textbf{Input Representation}} & \multicolumn{1}{c}{} \\
\cline{3-4}\cline{5-8}
\textbf{Work} & \textbf{Gran.} & \textbf{Model} & \makecell{\textbf{Design}\\\textbf{Strategy}} & \makecell{\textbf{Header}\\\textbf{Preprocessing}} & \makecell{\textbf{Header}\\\textbf{Fields Used}} & \makecell{\textbf{Input}\\\textbf{Shape}} & \makecell{\textbf{Input Size}\\\textbf{Selection}} & \makecell{\textbf{Hardware}\\\textbf{Aware}} \\
\hline
\cite{nas_etc_malekghaini2023automl4etc} & Packet & CNN/MLP & Search & IP anonym. & Fixed & 600B & Fixed & \xmark \\
\hline
\cite{lotfollahi2020deep} & Packet & 1D-CNN & Manual & \makecell{Eth. removal +\\IP anonym. + \\UDP zero-pad} & Fixed & 1500B & Fixed & \xmark \\
\hline
\cite{xu2024cascaded} & Flow & \makecell{Cascaded \\BLN }& Manual & \makecell{MAC/IP/port \\zero} & Fixed & 784B & Fixed & \xmark \\
\hline
\cite{nas_zhang2023automatic} & Flow & 1D-CNN & Search & \makecell{None\\reported} & Fixed & 784B & Fixed & \xmark \\
\hline
\cite{only_header_cui2022only} & Flow & CapsNet & Manual & IP zero & Fixed & 784B & Fixed & \xmark \\
\hline
\cite{yao2019identification} & Flow & Att.-LSTM & Manual & \makecell{Eth. removal +\\IP zero} & Fixed & $10\times1500$ & \makecell{Manual\\tuning} & \xmark \\
\hline
\cite{seydali2023cbs} & Hybrid & \makecell{1D-CNN + \\Bi-LSTM\\+ SAE} & Manual & \makecell{Eth. removal +\\IP anonym.} & Fixed & \makecell{1500B +\\25 feat.} & Fixed & \xmark \\
\hline
\cite{cui2019session} & Hybrid & CapsNet & Manual & MAC/IP zero & Fixed & 784B & Fixed & \xmark \\
\hline
\cite{wang2017malware} & Session & 1D-CNN & Manual & \makecell{MAC/IP \\anonym.} & Fixed & 784B & Fixed & \xmark \\
\hline
\cite{wang2017end} & Session & 1D-CNN & Manual & \makecell{MAC/IP \\anonym.} & Fixed & 784B & Fixed & \xmark \\
\hline
\cite{song2019encrypted} & Session & Text-CNN & Manual & Eth. removal & Fixed & 784B & Fixed & \xmark \\
\hline
\cite{wang2020encrypted} & Session & \makecell{1D-CNN +\\SAE} & Manual & \makecell{MAC/IP zero} & Fixed & 784B & Fixed & \xmark \\
\hline
\cite{zou2024novel_residual} & Session & \makecell{Residual \\CNN} & Manual & MAC/IP zero & Fixed & 784B & Fixed & \xmark \\
\hline
\cite{our_chehade2025iscc} & Session & 1D-CNN & Search & \makecell{Eth. removal +\\IP anonym.} & Fixed & 784B & Fixed & \cmark \\
\hline
\cite{our_chehade2025tnsm} & Session & 1D-CNN & Search & \makecell{Eth. removal +\\IP anonym.} & Fixed & 784B & Fixed & \cmark \\
\hline
\textbf{\makecell{Ours}} & \textbf{Session} & \textbf{1D-CNN} & \textbf{Search} & \textbf{Searchable} & \textbf{\makecell{Jointly\\optimized}} & \textbf{Searchable} & \textbf{\makecell{Jointly\\optimized}} & \textbf{\cmark} \\
\hline
\end{tabular}%
}
\end{table}

\section{Methodology}

RepTC is designed as a representation-aware, hardware-constrained search strategy that jointly optimizes network architecture and input representation to make compact neural models viable for TC on devices with even tens of kB of available RAM and milliwatt-range power consumption, while preserving satisfactory predictive performance. The methodology is organized into three stages: (i) the formulation of a resource-constrained optimization problem, (ii) the definition of a joint architecture/representation search space, and (iii) the application of an evolutionary search process to identify hardware-feasible, high-performing configurations. The next subsections elaborate on each stage.

\subsection{Joint Optimization Problem}


The hardware-constrained optimization problem jointly accounts for model architecture and input representation. Let \( \mathcal{A} \) be the set of candidate architectures, where \( a \in \mathcal{A} \) denotes a specific DNN; \( \mathcal{S}_h \) the set of header preprocessing strategies, with \( S_h \in \mathcal{S}_h \); and \( \mathcal{L}_s \) the set of admissible input lengths, with \( L_s \in \mathcal{L}_s \). The goal is to select a joint configuration \( (a, S_h, L_s) \) that maximizes validation accuracy while respecting tight resource constraints:

\vspace{-2mm}
\vspace{-2mm}
\begin{equation}
\begin{aligned}
\max_{a \in \mathcal{A},\, S_h \in \mathcal{S}_h,\, L_s \in \mathcal{L}_s}
\quad & \text{Accuracy}_{\text{val}}
\bigl(w^*(a,S_h,L_s), a, S_h, L_s\bigr) \\
\text{s.t.} \quad
& w^*(a,S_h,L_s) =
\argmin_w \mathcal{L}_{\text{train}}(w, a, S_h, L_s), \\
& |P(a)| < F_{\text{Th}}, \\
& |T(a,L_s)| < R_{\text{Th}}, \\
& \text{FLOPs}(a,L_s) < \text{FLOPs}_{\text{Th}} .
\end{aligned}
\label{eq:opt_hw_nas}
\end{equation}

Here, \( w^*(a,S_h,L_s) \) denotes the trained weights obtained by
minimizing the loss over the training set for architecture \( a \), given the
input configuration \( (S_h, L_s) \), which determines the input
representation.
The constraints ensure that the resulting configuration complies with edge platforms:

\begin{itemize}
    
\item \( |P(a)| \) is the total number of trainable parameters, which determines the required Flash storage. It depends only on \( a \) because parameters are fixed by the model structure.

\item \( |T(a, L_s)| \) is the number of elements in the largest activation tensor during inference; this parameter affects peak RAM usage, and depends on both \( a \) and \( L_s \), since tensor dimensions grow with layer shape and input size.

\item  \( \text{FLOPs}(a, L_s) \) is the total number of floating-point operations per inference, affected by computing and energy costs. It also depends on both \( a \) and \( L_s \), as more complex architectures and longer inputs require more operations.

\end{itemize}

The scalar quantities \( F_{\text{Th}} \), \( R_{\text{Th}} \), and \( \text{FLOPs}_{\text{Th}} \) set thresholds on Flash, RAM, and compute, respectively. These constraints stem from previous work on the support of inference on microcontroller-class devices~\cite{ragusa2024combining, ragusa2024compression, our_chehade2025iscc}. The formulation in \eqref{eq:opt_hw_nas} enables RepTC to approximate a hardware-feasible near-optimal configuration in a multidimensional search space; within this space, architectural and input variables interact through the hardware constraints.

\subsection{Configuration of Input Representation}

We choose to operate on raw traffic data to avoid the delay and complexity of manual feature extraction, as deep networks can automatically learn relevant features from raw bytes; the analysis is therefore conducted at the session level, which captures bidirectional exchanges and exposes higher-level dependencies compared to individual packets or unidirectional flows. 
In this setting, sessions are formed by grouping packets belonging to the same bidirectional communication, identified by shared source/destination IPs and protocol fields (e.g., TCP/UDP ports, ICMP type/code, or ARP headers).
The resulting byte sequence preserves both headers and payload content. The input length, \( L_s \), specifies how many bytes are processed per session and therefore sets the model’s input size. 
RepTC treats \( L_s \in \mathcal{L}_s \) as a discrete search variable that trades off accuracy with memory and computational cost.

Each session includes multiple packets with metadata across protocol layers, ranging from Ethernet (e.g., MAC address, EtherType) to IP (e.g., source/destination IPs) and transport-layer fields (e.g., TCP/UDP ports). 
These fields are commonly transformed during preprocessing to limit overfitting to static identifiers and enhance privacy, while preserving structural patterns relevant for classification. Techniques include zeroing (e.g., replacing field values with zeros), anonymization (e.g., replacing identifiers with synthetic values), field removal, and padding-based alignment. Because their impact depends on both model behavior and input structure, RepTC treats header preprocessing as a categorical search dimension.

Therefore, the set \( \mathcal{S}_h \) spans eight preprocessing strategies, combining operations such as Ethernet removal (Eth. Rem.), MAC zeroing (MAC Zero), IP anonymization (IP Anon.), IP zeroing (IP Zero), and port zeroing (Port Zero). Table~\ref{tab:preprocessing_strategies} lists the strategies in \( \mathcal{S}_h \), where each column represents a strategy \( S_h \), and each row indicates the application of a specific operation (\checkmark) to the corresponding header field. 
The selection excludes operations that introduce protocol-specific assumptions. In particular, transformations that depend on specific protocol structures (e.g., UDP padding) were not considered, as they may not generalize to heterogeneous traffic streams mixing ARP, ICMP, and application-layer protocols. As a result, \( \mathcal{S}_h \) retains a representative set of widely used, protocol-agnostic operations, keeping the search space both relevant and tractable.

Input representation thus becomes an adaptive axis: both \( S_h \in \mathcal{S}_h \) (categorical) and \( L_s \in \mathcal{L}_s \) (discrete) are integrated into the search space and co-optimized with architecture to match data structure and scale.

\begin{table}[ht]
\centering
\caption{Preprocessing strategies defined in \( \mathcal{S}_h \)}
\label{tab:preprocessing_strategies}
\renewcommand{\arraystretch}{1.1}
\scriptsize
\resizebox{\columnwidth}{!}{%
\begin{tabular}{l|c|c|c|c|c|c|c|c}
\hline
\textbf{Operation} & \textbf{S1} & \textbf{S2} & \textbf{S3} & \textbf{S4} & \textbf{S5} & \textbf{S6} & \textbf{S7} & \textbf{S8} \\
\hline
Eth. Rem.   & \centering\checkmark & \checkmark & \checkmark & \checkmark & & & & \\
\hline
MAC Zero    & & & & & \checkmark & \checkmark & \checkmark & \checkmark \\
\hline
IP Anon.    & \checkmark & \checkmark & & & \checkmark & \checkmark & & \\
\hline
IP Zero     & & & \checkmark & \checkmark & & & \checkmark & \checkmark \\
\hline
Port Zero   & \checkmark & & \checkmark & & \checkmark & & \checkmark & \\
\hline
\end{tabular}
}
\end{table}


\subsection{Architecture Search Space}

The search space \( \mathcal{A} \) \added{used in the main experiments} focuses on lightweight 1D-CNNs, since previous research \cite{wang2017end, wang2017malware, lotfollahi2020deep, our_chehade2024tiny} shows that these architectures can attain notable accuracy on raw traffic. Moreover, 1D-CNNs are more amenable to hardware-aware optimization than RNNs or Transformers, as their localized operations and weight sharing enable efficient implementations with low computational and memory overhead on resource-constrained devices\cite{tay2022efficient, cnn_varughese2025optimization}. \added{Section~\ref{sec:architecture_transfer} further evaluates RepTC with a
non-convolutional architecture space.}

Each candidate model includes a stack of modular blocks, and the search process also adjusts the number of blocks. A block includes a 1D convolutional layer followed by batch normalization, ReLU activation, optional pooling (max or average), and dropout. The RepTC process tunes the related hyperparameters (i.e., kernel size, filter count, stride, pooling type, pool size, and padding type) to control architectural capacity and receptive field. The topmost stages in each architecture include a global average pooling (GAP) layer and a dense classifier. This compact structure preserves relevant session-level features while minimizing inference cost.

\subsection{Joint Evolutionary Search}\label{sec:evolution}

The optimization process requires an effective method to explore the search space under hardware constraints. Toward that end, an evolutionary method optimizes the architecture \( a \in \mathcal{A} \), the preprocessing \( S_h \in \mathcal{S}_h \), and the input size \( L_s \in \mathcal{L}_s \) jointly. Evolutionary approaches are well suited to these types of mixed-variable optimization problems, and have proven effective in related domains~\cite{1000papersNAS, ragusa2024combining, ragusa2024compression}. In the following, \( \mathcal{X_T} \) and \( \mathcal{X_V} \) denote the training and validation sets, respectively, and \( \mathcal{H} \) represents the hardware constraints that define thresholds on parameters, intermediate tensor size, and FLOPs.

The process starts from an initial parent configuration, \( (a_0, S_h^0, L_s^0) \), and runs for a fixed budget of \(N_g\) generations. 
At each generation, \(N_c\) child configurations are derived from the current parent through a mutation operator \( R_m(a_p, S_h^p, L_s^p) \). 
Architectural mutations can add, remove, or modify blocks. These actions are selected randomly: adding inserts a new convolutional block into the sequence; removing deletes one (while keeping at least a single block); and modifying perturbs a block by randomly resampling its hyperparameters (e.g., filter count, kernel size, stride) within the same architecture space. Add-block mutations are bounded by a maximum depth to limit model growth and keep the search tractable. Each child undergoes a fixed number of such architectural mutations.

On the input side, \( S_h \) mutates by either keeping the current configuration or switching to another in \( \mathcal{S}_h \setminus \{S_h^p\} \). 
Mutations in the input size evolve by either keeping the parent value, \( L_s^c = L_s^p \), or updating it as \( L_s^c = L_s^p + \Delta L_s \), where \( \Delta L_s \) is sampled uniformly from \( [-\Delta_{\text{max}}, +\Delta_{\text{max}}] \).
Reflective clipping enforces bounds on \( L_s^c \) by mirroring excess values at range limits.

\

Each candidate is evaluated against hardware constraints \( \mathcal{H} \), and those that exceed any threshold are discarded. Mutations continue until the required number of admissible children \(N_c\) per generation is reached.
The resulting HW-admissible children undergo the conventional training/validation procedure. 
Selection follows a best-offspring replacement strategy among the \(N_c\)
admissible children, in which the child yielding the highest validation
accuracy as measured by \(E(a,S_h,L_s,\mathcal{X_V})\) becomes the new
parent \( (a_p,S_h^p,L_s^p) \) for the next generation.
The process tracks the best-so-far configuration \( (a^*, S_h^*, L_s^*) \) across generations and returns it at the end. The procedure is summarized in Algorithm~\ref{alg:hw_nas}.

\begin{algorithm}
    \caption{RepTC: Joint Search over Architecture, Preprocessing, and Input Size}
    \label{alg:hw_nas}

    \textbf{Inputs:} Training set \( \mathcal{X_T} \), validation set
    \( \mathcal{X_V} \); candidate architectures \( \mathcal{A} \);
    preprocessing strategies \( \mathcal{S}_h \); input sizes
    \( \mathcal{L}_s \); initial parent \( (a_0, S_h^0, L_s^0) \);
    mutation operator \( R_m \); generations \( N_g \); admissible children
    per generation \(N_c\); evaluation function
    \( E(\cdot,\mathcal{X_V}) \); hardware constraints \( \mathcal{H} \).

    \begin{algorithmic}
      \State \textbf{0. Init:} Set parent
      \( (a_p, S_h^p, L_s^p) = (a_0, S_h^0, L_s^0) \)

      \State Set best-so-far
      \( (a^*, S_h^*, L_s^*) \leftarrow (a_p, S_h^p, L_s^p) \)

      \For{$g = 1, \dots, N_g$}
          \Comment{Iterate through generations}

          \For{$c = 1, \dots, N_c$}
              \Comment{Generate $N_c$ children}

              \State
              \( (a_c, S_h^c, L_s^c)
              \gets R_m(a_p, S_h^p, L_s^p) \)
              \Comment{Apply mutations}

              \If{\((a_c, L_s^c)\) satisfies \(\mathcal{H}\)}
                  \State Train
                  \( (a_c, S_h^c, L_s^c) \)
                  on \( \mathcal{X_T} \)

                  \State Evaluate using
                  \( E(a_c, S_h^c, L_s^c, \mathcal{X_V}) \)
              \EndIf
          \EndFor

          \State \textbf{Selection:} Update parent with the best valid child

          \Statex
          \[
          (a_p, S_h^p, L_s^p)
          =
          \argmax_{c = 1, \dots, N_c}
          E(a_c, S_h^c, L_s^c, \mathcal{X_V})
          \]

          \State If
          \(E(a_p,S_h^p,L_s^p,\mathcal{X_V})
          >
          E(a^*,S_h^*,L_s^*,\mathcal{X_V})\),
          set
          \( (a^*, S_h^*, L_s^*)
          \leftarrow
          (a_p, S_h^p, L_s^p) \)

      \EndFor

      \State \textbf{Output:} Optimal configuration
      \( (a^*, S_h^*, L_s^*) \)

    \end{algorithmic}
\end{algorithm}


\section{Experimental Setup}

\subsection{Datasets}

Three real-world benchmarks support the experimental evaluation, covering a variety of critical domains in network-traffic analysis. All datasets include raw traffic in PCAP format, enabling session-level segmentation and processing.

\textbf{ISCX VPN-nonVPN}~\cite{iscxvpnnonvpn} holds real-world heterogeneous traffic, captured under VPN and non-VPN conditions. It has been widely adopted in ETC research~\cite{wang2017end, wang2017malware, lu2021iclstm, zou2024novel_residual}. \textbf{USTC-TFC2016}~\cite{wang2017malware} contains traffic from ten benign apps and ten malware families, mixing encrypted and unencrypted samples. \textbf{Edge-IIoTset}~\cite{edgeiiot_ferrag2022edge} covers Industrial IoT (IIoT) scenarios, with both normal sensor traffic and various attacks. Its IIoT-specific protocols and structures allow evaluating the method's effectiveness under lightweight intrusion detection constraints. 

\begin{figure}[htbp]
    \centering
    \includegraphics[width=\columnwidth]{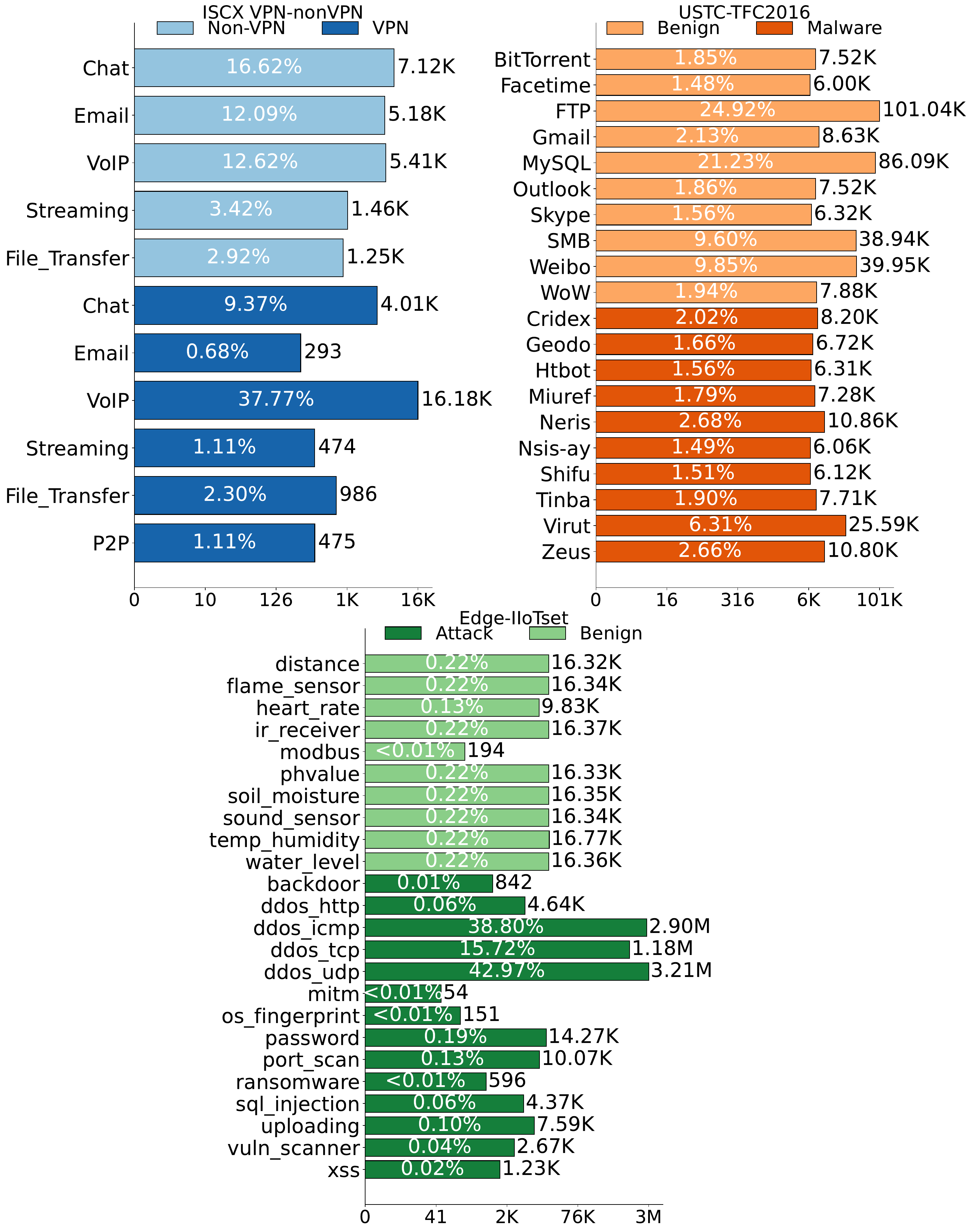}
    \caption{Class distribution overview for all evaluation datasets.}
    \label{fig:dataset_distribution}
\end{figure}

Figure~\ref{fig:dataset_distribution} gives the distributions of classes (sample count and relative percentage) in each benchmark. In the top-left and top-right panels, (ISCX and USTC, respectively) colors distinguish VPN/non-VPN and benign/malware traffic. All datasets exhibit some degree of class imbalance; however, Edge-IIoTset is particularly skewed, as three Distributed Denial-of-Service (DDoS) attack classes (\textit{ICMP}, \textit{TCP SYN}, and \textit{UDP flood}) contain millions of samples, far exceeding the others.

\subsection{RepTC Setup and Implementation} \label{setup}

The implementation of RepTC to perform tests on the datasets relied on TensorFlow and Keras, and all experiments ran on a workstation hosting an NVIDIA 3070 Ti GPU. 

Session extraction and preprocessing used Scapy, where packets from the same bidirectional session were aggregated into a single input sequence; duplicate session samples within the same class were removed to avoid redundancy \cite{nas_zhang2023automatic}.
Header fields were processed according to the strategies \(S_h\) listed in Table~\ref{tab:preprocessing_strategies}, with identifying information consistently transformed. In compliance with the work in \cite{wang2017end, wang2017malware}, sessions were represented with an initial length of \( L_s = 784 \) bytes, applying zero-padding for shorter sessions and truncation for longer ones. For each \(S_h\), a preprocessed dataset variant was generated. Byte values were normalized by dividing by 255 to map inputs to \([0,1]\). These variants served as selectable inputs during the RepTC search.
For each dataset, the resulting session samples were partitioned into training and held-out test subsets (80/20 split) while preserving class proportions, ensuring that no session (or any of its packets) is shared across splits. During the search procedure, 20\% of the training subset was reserved for validation.
In the Edge-IIoTset dataset, the three DDoS-dominated classes (\textit{ICMP}, \textit{TCP SYN}, and \textit{UDP flood}) were randomly undersampled to 30\,000 samples each within the training and validation subset to mitigate imbalance, whereas the test distribution was left unchanged.

\begin{figure}[htbp]
    \centering
    \includegraphics[width=\columnwidth]{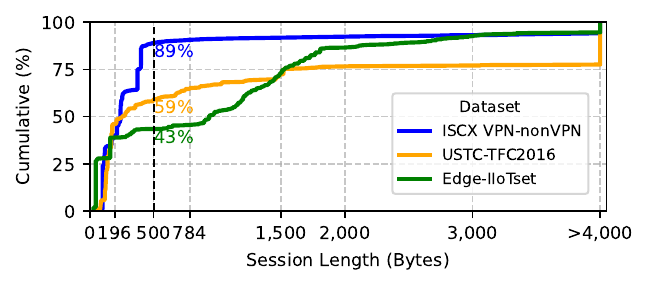}
    \caption{Cumulative session length distribution across datasets.}

    \label{fig:session_length_kde}
\end{figure}

Figure~\ref{fig:session_length_kde} shows the cumulative session-length distributions across all extracted sessions for the three datasets. The x-axis gives the session length (in bytes), while the y-axis reports on the percentage of sessions having length up to the coordinate value. The graph indicates that the initial 500 bytes cover between 43\% and 89\% of sessions, depending on the dataset.

As a consequence, the input length \( L_s \) in bytes was defined as a searchable variable in the range \([196, 784]\), to balance efficiency and early-packet information retention. The lower bound (196 bytes) was set to capture at least the initial packet(s); the upper bound (784 bytes) was chosen to remain consistent with widely used session-based ETC settings while still allowing the search to explore shorter, more efficient alternatives. Mutations in the input length involved even-valued steps drawn from 
\(\Delta L_s \sim \mathrm{Uniform}(\{-128, -126, \dots, 126, 128\})\), with symmetric reflection at the boundaries (i.e., 196 and 784) to maintain search space validity, as described in Section~\ref{sec:evolution}.

When defining the hardware thresholds, prior session-level models were examined through architecture-level complexity metrics reported in the literature, complemented, when possible, by offline profiling or Keras-based reconstruction to
estimate parameter count, maximum intermediate tensor size,
and FLOPs. These metrics served as proxies for model storage, activation memory, and computational cost, and revealed significant headroom relative to the deployment budgets.


The following constraints in the search process ensure compatibility with a broad range of lightweight IoT platforms available on the market: parameter count \(< 60\text{k}\) (240kB Flash), maximum intermediate tensor size \(< 4\text{k}\) elements (16kB RAM), and inference cost \(< 500\text{k}\) FLOPs. These thresholds comply with platforms hosting at least 256kB Flash and 32kB RAM, and extend to devices with larger budgets. \superadded{They thus target a broad class of MCU-scale devices spanning heterogeneous processor architectures.} For the computational constraint, \deleted{Cortex-M cores, which dominate IoT-class microcontrollers, are theoretically capable of sustaining approximately one arithmetic operation per clock cycle~\cite{lai2018cmsis}; thus a 500k-FLOP model maps to execution times on the order of 3--7\,ms at typical operating frequencies (80--200\,MHz), before accounting for memory stalls or communication overheads;}\superadded{Cortex-M cores provide a representative MCU reference and are theoretically capable of sustaining approximately one arithmetic operation per clock cycle~\cite{lai2018cmsis}; thus a 500k-FLOP model maps to execution times on the order of 3--7\,ms at typical operating frequencies (80--200\,MHz), before accounting for memory stalls and architecture-dependent implementation overheads.} Cortex-A gateways and dedicated NPUs (Neural Processing Units) or TPUs (Tensor Processing Units) achieve much lower latencies thanks to higher frequencies and specialized compute units.
These estimates still provide ample headroom with respect to the (100--250\,ms) latency range reported as acceptable for embedded monitoring tasks~\cite{jiang2018low_flops_latency}. Empirical deployment results in Section~\ref{deployment}, obtained after quantization, confirm that actual inference latencies of the final RepTC models remain well below this range.

The search spanned 100 generations, with 20 candidate architectures per generation; each search evaluated 2000 admissible candidates per dataset, rejecting infeasible ones before training under the hardware constraints. The wall-clock time for the RepTC search was about 18--22\,h per dataset. 

Networks included 1D convolutional blocks, with filters in [16, 140], kernel sizes in [1, 7], strides in [1, 7], and optional max or average pooling with pool sizes of 2 or 3. Padding options were “same” and “valid”. Architectural depth was capped at four convolutional blocks, and dropout rates were scaled linearly with depth (from 0.1 to 0.5) to regularize deeper models.
For each child, up to two block-level mutations were applied to induce sufficient structural variation. 
In addition to architectural changes, the preprocessing strategy \(S_h\) and the input length \(L_s\) were mutated independently for each child, each with probability 0.5.


\deleted{The training of each candidate involved 200 epochs for ISCX and 30 epochs for USTC and Edge-IIoTset,}
\added{Each candidate was trained with a maximum budget of 200 epochs for ISCX and 30 epochs for USTC and Edge-IIoTset,}
relying on the Adam optimizer (initial learning rate \(10^{-3}\), batch size 256), and categorical cross-entropy loss, with adaptive decay via \textit{ReduceLROnPlateau} and early stopping based on validation loss.

\added{Table~\ref{tab:training_budget} summarizes the candidate-training budgets under the batch size of 256 used for all datasets. Since the training-set sizes differ substantially, an epoch corresponds to a different number of optimization steps in each case (i.e., the candidate-training samples processed in batches of 256). USTC-TFC2016 and Edge-IIoTset therefore involve considerably more updates per epoch than ISCX VPN-nonVPN. Their shorter maximum epoch counts keep the overall training budgets within a comparable range while limiting the training cost per candidate. Early stopping may terminate training before these maximum budgets are reached when validation loss no longer improves.}

{\color{blue}
\begin{table}[t]
\centering
\caption{\added{Candidate-training budgets across datasets.}}
\label{tab:training_budget}

\renewcommand{\arraystretch}{0.95}
\small

\begin{adjustbox}{max width=\columnwidth}
\begin{tabular}{
>{\centering\arraybackslash}c|
>{\centering\arraybackslash}c|
>{\centering\arraybackslash}c|
>{\centering\arraybackslash}c|
>{\centering\arraybackslash}c
}
\hline

\textbf{Dataset}
& \makecell{\textbf{Candidate}\\\textbf{Training Samples}}
& \makecell{\textbf{Batches}\\\textbf{/ Epoch}}
& \makecell{\textbf{Maximum}\\\textbf{Epochs}}
& \makecell{\textbf{Maximum}\\\textbf{Updates}}
\\
\hline

ISCX VPN-nonVPN
& 27,426
& 108
& 200
& 21,600
\\
\hline

USTC-TFC2016
& 259,530
& 1,014
& 30
& 30,420
\\
\hline

Edge-IIoTset
& 172,907
& 676
& 30
& 20,280
\\
\hline

\end{tabular}
\end{adjustbox}
\end{table}
}

The best configuration found by RepTC, i.e. the one achieving the highest validation accuracy across generations for each dataset, was then trained on the full training subset and evaluated on the held-out test subset; this procedure was repeated over ten independent runs, and the corresponding sample standard deviation was reported.

\section{Results} \label{results}

\subsection{RepTC Architectures and Resource Profiles}

Table~\ref{tab:cnn_architectures_combined} reports the best architectures found by RepTC for each dataset. Each block shows one model, with rows listing layers and columns indicating layer type, input shape, number of filters or units (Filt/Units), kernel or pool size (Ker/Pool), stride (Str), and padding (Pad). The header above each model summarizes its profile: number of parameters (Params), maximum intermediate tensor size (M-Tens), FLOPs, selected preprocessing strategy \( S_h \), and input length \( L_s \).

\begin{table}[htbp]
\centering
\caption{Best architectures identified by RepTC }
\label{tab:cnn_architectures_combined}
\renewcommand{\arraystretch}{1.1}
\resizebox{\columnwidth}{!}{%
\begin{tabular}{c|c|c|c|c|c|c}
\hline
\textbf{\#} & \textbf{Layer} & \textbf{Input Dim} & \textbf{Filt/Units} & \textbf{Ker/Pool} & \textbf{Str} & \textbf{Pad} \\
\hline
\multicolumn{7}{c}{\textbf{RepTC-ISCX} model (ISCX VPN-NonVPN dataset)} \\
\multicolumn{7}{c}{\textbf{Params:} \textbf{54.67k}, \textbf{M-Tens:} \textbf{2.55k}, \textbf{FLOPs:} \textbf{305.87k}, \( \mathbf{S_h = 4} \), \( \mathbf{L_s = 218} \)} \\
\hline
1 & Conv1D + ReLU   & 218 x 1   & 69  & 7 & 6 & same \\
2 & Conv1D + ReLU   & 37 x 69   & 61  & 4 & 7 & valid \\
3 & Conv1D + ReLU   & 5 x 61    & 48  & 1 & 1 & same \\
4 & Conv1D + ReLU   & 5 x 48    & 132 & 5 & 5 & same \\
5 & GAP             & 1 x 132   & -   & - & - & - \\
6 & Dense + Softmax & 1 x 132   & 11  & - & - & - \\
\hline
\multicolumn{7}{c}{\textbf{RepTC-USTC} model (USTC2016 dataset)} \\
\multicolumn{7}{c}{\textbf{Params:} \textbf{28.84k}, \textbf{M-Tens:} \textbf{1.52k}, \textbf{FLOPs:} \textbf{482.96k}, \( \mathbf{S_h = 8} \), \( \mathbf{L_s = 530} \)} \\
\hline
1 & Conv1D + ReLU   & 530 x 1   & 20  & 7 & 7 & same \\
2 & Conv1D + ReLU   & 76 x 20   & 53  & 7 & 4 & valid \\
3 & Max Pooling     & 18 x 53   & -   & 2 & 2 & valid \\
4 & Conv1D + ReLU   & 9 x 53    & 114 & 3 & 2 & same \\
5 & GAP             & 1 x 114   & -   & - & - & - \\
6 & Dense + Softmax & 1 x 114   & 20  & - & - & - \\
\hline
\multicolumn{7}{c}{\textbf{RepTC-Edge} model (Edge-IIoTset dataset)} \\
\multicolumn{7}{c}{\textbf{Params:} \textbf{25.34k}, \textbf{M-Tens:} \textbf{1.14k}, \textbf{FLOPs:} \textbf{262.44k}, \( \mathbf{S_h = 5} \), \( \mathbf{L_s = 314} \)} \\
\hline
1 & Conv1D + ReLU   & 314 x 1   & 22  & 5 & 6 & valid \\
2 & Conv1D + ReLU   & 52 x 22   & 59  & 6 & 6 & valid \\
3 & Conv1D + ReLU   & 8 x 59    & 83  & 3 & 2 & same \\
4 & GAP             & 4 x 83    & -   & - & - & - \\
5 & Dense + Softmax & 1 x 83    & 24  & - & - & - \\
\hline
\end{tabular}%
}
\end{table}

The resulting, compact architectures all satisfied the design constraints for each benchmark, and could easily be deployed on sub-256kB Flash and 32kB RAM platforms. 
The architecture identified for \textit{RepTC-ISCX} relies on aggressive downsampling through large strides (up to 7), enabling deeper convolutional stacking without pooling. The \textit{RepTC-USTC} model instead follows a progressive increase in filter count and includes a pooling stage to control intermediate tensor growth. For \textit{RepTC-Edge}, the search converged to a minimal three-layer configuration with gradually increasing filter capacity.

The eventual input length varied across datasets. This reflected some adaptation to session scale, yielding shorter inputs for \textit{ISCX} and \textit{Edge-IIoTset} (\( L_s = 218 \) and \( L_s = 314 \), respectively) and longer ones for \textit{USTC} (\( L_s = 530 \)), mostly to capture extended traffic patterns. 
This variation suggests that input length should be treated as a tunable design variable rather than fixed a priori to the conventional 784-byte setting, since lower values can remain effective.
Likewise, the diversity in selected preprocessing strategies (\( S_h = 4, 8, 5 \)) indicates that effective header handling is context-dependent in practice.


\subsection{Analysis of Search Space Factors}\label{sec:statistical}

Understanding which elements of the search space are most strongly
associated with performance helps interpret the design outcomes and guide
future practitioners. This section quantifies how architectural and input-level
variables relate to model performance, providing statistical evidence for the
joint optimization strategy.
A total of 2000 configurations evaluated during the search for each dataset (100 generations with 20 offspring each) were analyzed based on their population statistics. 
The analysis relied on mutual information (MI), Pearson correlation, and multifactor ANOVA to quantify how each variable is associated with validation accuracy.
MI was estimated using a non-parametric $k$-nearest-neighbor approach ($k=5$)~\cite{kraskov2004estimating}, suitable for mixed discrete and categorical factors. 
It measures how much information each variable provides
about differences in accuracy, independently of direction.
Pearson correlation indicates whether the association is positive or negative.
ANOVA tested whether variations in factor values (or their interactions) corresponded to statistically significant changes in accuracy across the explored architectures.  
For variables appearing in multiple layers (e.g., kernel size, stride, filter count), average values were computed per configuration before analysis to obtain consistent descriptors.

\begin{figure}[htbp]
    \centering
    \includegraphics[width=\columnwidth]{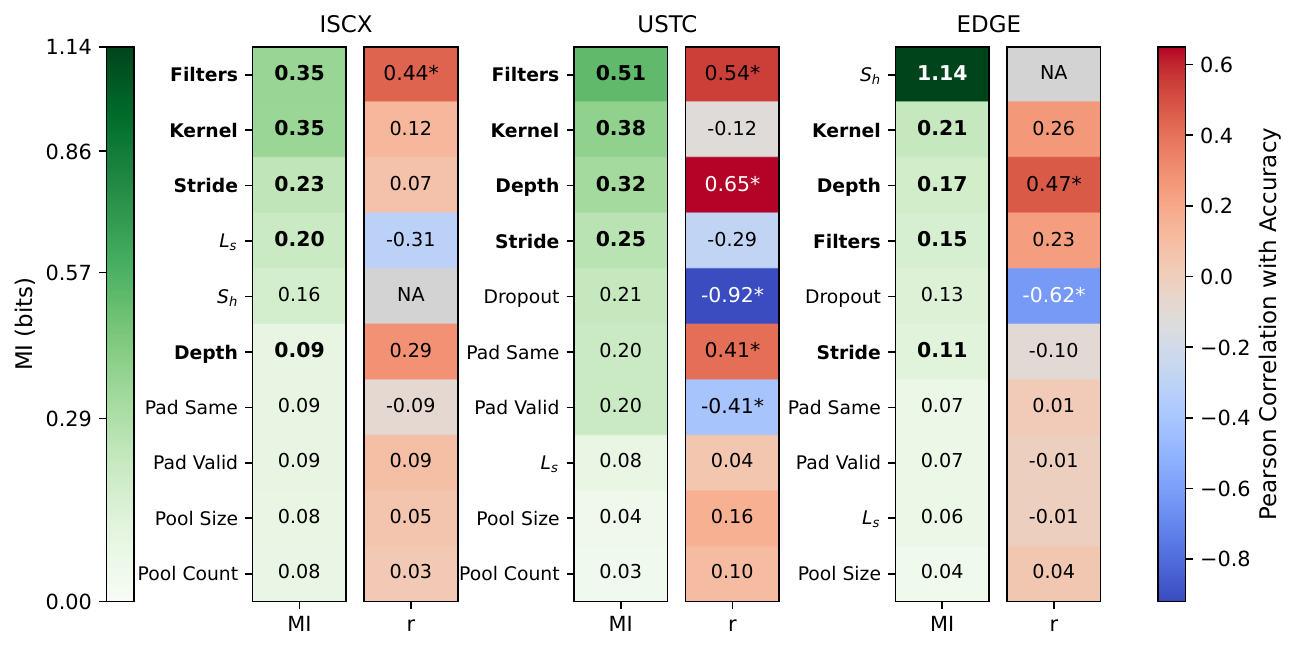}
    \caption{Mutual information and Pearson correlation of search-space factors with validation accuracy.}
    \label{fig:mi_corr_heatmaps}
\end{figure}

Figure~\ref{fig:mi_corr_heatmaps} shows the top ten factors per dataset by MI and their correlation values.
For each dataset pair (left: MI, right: correlation), rows denote architectural or input variables; 
green intensity indicates higher MI, 
while blue and red in the correlation ($r$) maps indicate negative and positive associations with accuracy, respectively. Rows in bold mark the top-4 MI factors or global stars, i.e., factors ranked in the top four in at least two datasets; 
cells in the correlation maps associated with the categorical preprocessing
strategy \(S_h\) are shown in grey and annotated ``NA,'' since Pearson
correlation is not meaningful under arbitrary numerical encoding;
and an asterisk flags $|r| \geq 0.40$, denoting practically relevant effects.

On \textit{ISCX}, kernel and filter size both rank highest (MI = 0.35\,bits), followed by stride (0.23), input length \(L_s\) (0.20), and preprocessing strategy \(S_h\) (0.16), while depth has moderate importance. 
Accuracy increases with filter count ($r=0.44$) and slightly with depth ($r=0.29$); the negative correlation for \(L_s\) ($r=-0.31$) indicates that smaller inputs tend to achieve higher accuracy while reducing resource usage, 
supporting input size as a key optimization variable in the joint search.
For \textit{USTC}, filter and kernel size remain dominant (0.51 and 0.38\,bits), followed by depth (0.32), which is more relevant here; their positive correlations ($r=0.54$ and $r=0.65$) indicate that accuracy scales with model capacity in this dataset, while input factors contribute less. 
On \textit{Edge-IIoTset}, preprocessing (\(S_h\)) is the dominant factor (MI = 1.14\,bits), showing that header configuration is strongly associated with accuracy, whereas kernel and depth have secondary effects (0.21 and 0.17\,bits). 
Together, the MI and correlation analyses indicate that receptive-field and filter tuning are the most influential variables, while input length and preprocessing contribute in a complementary way and, in some cases such as \(S_h\) in the Edge dataset, become primary factors. \added{The origin of this cross-dataset difference is examined at the class level in Section~\ref{input_sensitivity}.} Parameters such as pooling and padding show weaker associations with accuracy, but still play a role in feature mapping and dimensionality control.

\begin{figure}[htbp]
    \centering
    \includegraphics[width=\columnwidth]{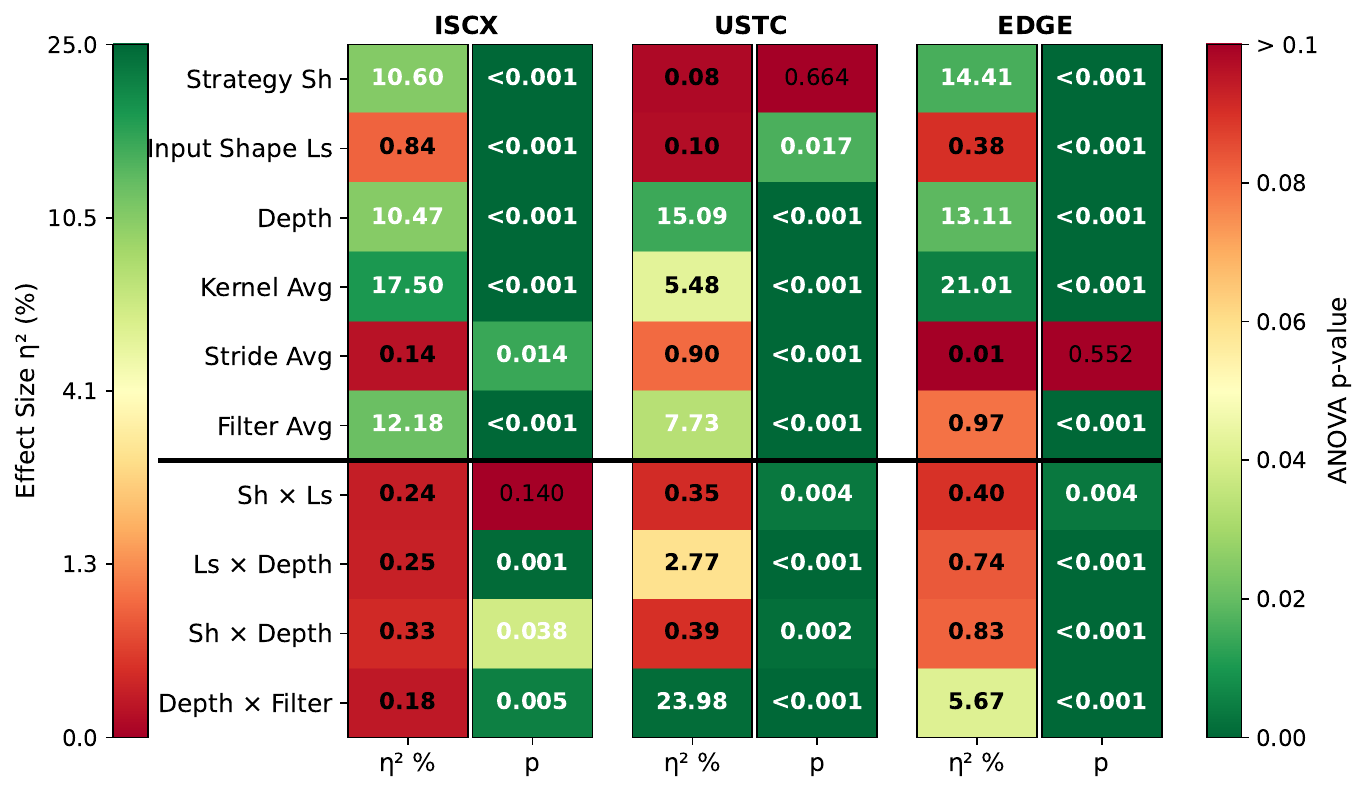}
    \caption{ANOVA $p$-value and $\eta^2$ effect-size heatmaps for key architectural and input factors.}

    \label{fig:anova_heatmaps}
\end{figure}

Figure~\ref{fig:anova_heatmaps} summarizes representative ANOVA outcomes for the principal architectural variables (depth, kernel, stride, and filters), the input-related factors (\(S_h\), \(L_s\)), and selected pairwise interactions (e.g., Depth $\times$ Filter, \(S_h \times L_s\), and \(L_s \times\)Depth).
A horizontal black divider separates factors (above) from interaction terms (below). 
Each dataset includes paired heatmaps showing the effect size ($\eta^2$) on the left and the corresponding $p$-values on the right. 
$\eta^2$ represents the percentage of variance in validation accuracy explained by each factor or interaction; higher values (green shades, up to 25\%) indicate stronger effects, 
while lower values appear in red, denoting weaker influence. 
The $p$-values denote the statistical significance of each effect (Benjamini–Hochberg corrected); lower values (dark green, below 0.001) denote high confidence in the observed difference, whereas values approaching or exceeding 0.1 indicate limited significance. 
Residual variance ranges from 31\% to 42\%; thus, the analyzed components explain about two-thirds of the variability in model accuracy. 

Across datasets, kernel size and depth emerge as influential architectural
factors, while filter count also shows substantial effects on ISCX and USTC
but a smaller effect on Edge-IIoTset ($\eta^2=0.97\%$).
The Depth $\times$ Filter interaction ($\eta^2$ up to approximately 24.0\%)
further suggests that receptive-field control and model capacity are central
to performance. Stride contributes little, except for \textit{USTC}, where it shows a mild effect. On the input side, the preprocessing strategy \(S_h\) is particularly relevant for both \textit{ISCX} and \textit{Edge-IIoTset} ($\eta^2$ = 10.6\% and 14.4\%, respectively), showing a substantial association between header processing and
validation accuracy in these datasets. The input shape \(L_s\) is also consistently significant, with moderate effect size, indicating that the analyzed session length affects accuracy as well as computational cost. The interaction \(S_h\times L_s\) is significant in both \textit{USTC} and \textit{Edge-IIoTset} (low $\eta^2$ but $p<0.01$); in \textit{USTC}, preprocessing becomes relevant only when coupled with input length, while in \textit{Edge-IIoTset} the combination slightly strengthens the effect already evident for \(S_h\). Other input–architecture interactions, such as \(L_s\times\)Depth and \(S_h\times\)Depth, are statistically significant but have low effect size; these interactions still collectively contribute to the accuracy improvements observed across models.

Overall, the ANOVA findings complement the MI and correlation analyses by clarifying how these factors interact within the search space. 
The results show that architectural and input-representation factors
are jointly associated with the observed performance differences, further
supporting the integrated search-space formulation.

\subsection{Performance Comparison with Existing Methods}

The first row in Table~\ref{tab:sota_summary} lists in boldface the outcomes of the RepTC strategy; each subsequent row corresponds to a representative method reported in the literature. The second column indicates the input granularity: \textit{Sess.}\ denotes raw session bytes; \textit{Flows/Pack.}\ refers to flow- or packet-level bytes; \textit{Hybrid}\ combines multiple traffic granularities; and \textit{Features}\ uses pre-computed statistical flow features. The remaining columns report predictive and architectural metrics: accuracy (Acc.), weighted F1-score (F1\textsubscript{w}), macro F1-score (F1\textsubscript{m}), parameter count (Params., in millions), peak tensor size (M-Tens., in kiloelements), and inference cost (FLOPs, in millions). Accuracy and F1 values are taken from the original studies, whereas the architectural metrics were obtained from the papers when reported or otherwise recomputed from the published model descriptions under a consistent Keras-based profiling procedure. Params., M-Tens., and FLOPs are used as architecture-level indicators of model size, peak activation memory, and computational cost. A double horizontal line separates session-based methods from approaches using other input granularities, since raw-session modeling is the primary focus of this study; comparisons within the session-based group are therefore the most direct, whereas the remaining methods provide broader context because different granularities expose different traffic patterns, require different preprocessing pipelines, and may lead to different inference-trigger rates in real-time operation. For each dataset block, values are color-coded: green cells mark favorable values (high Acc./F1 or lower resource usage), whereas red cells highlight weaker performances. Results of the proposed models are reported as mean $\pm$ standard deviation over ten independent runs.

\begin{table}[htbp]
    \centering
  \caption{Comparison with representative prior works in terms of reported predictive performance and architecture-level complexity indicators.}

    \label{tab:sota_summary}
    \renewcommand{\arraystretch}{0.981}
    \large
    \resizebox{\columnwidth}{!}{%
    \begin{tabular}{cccccccc}
        \hline
        \makecell{\textbf{Method}} &
        \makecell{\textbf{Input}} &
        \makecell{\textbf{Acc.} \\ (\%)} &
        \makecell{\textbf{F1\textsubscript{W}} \\(\%)} &
        \makecell{\textbf{F1\textsubscript{m}} \\(\%)} &
        \makecell{\textbf{Params.} \\ (M)} &
        \makecell{\textbf{M-Tens.} \\ (K)} &
        \makecell{\textbf{FLOPs} \\ (M)} \\
        \hline
        & & & & & & & \\[-0.535cm]
\multicolumn{8}{c}{\textbf{ISCX VPNNon-VPN Dataset}} \\
\hline
\textbf{\textbf{RepTC-ISCX}} & \textbf{\textbf{sess}.} & \cellcolor[rgb]{0.55,0.81,0.40}\textbf{\makecell{96.21\\ $\pm$\textbf{0.51}} } & \cellcolor[rgb]{0.62,0.84,0.41}\textbf{\makecell{96.24\\ $\pm$\textbf{0.51}} } & \cellcolor[rgb]{0.65,0.00,0.15}\textbf{\makecell{95.33\\ $\pm$\textbf{0.48}} } & \cellcolor[rgb]{0.00,0.41,0.22}\textbf{0.05} & \cellcolor[rgb]{0.14,0.62,0.32}\textbf{2.55} & \cellcolor[rgb]{0.00,0.41,0.22}\textbf{0.30} \\
\cite{our_chehade2025tnsm} & Sess. & \cellcolor[rgb]{0.47,0.77,0.40}96.60 & \cellcolor[rgb]{0.54,0.80,0.40}96.63 & - & \cellcolor[rgb]{0.00,0.42,0.22}0.09 & \cellcolor[rgb]{0.73,0.88,0.46}20.12 & \cellcolor[rgb]{0.57,0.81,0.41}10.08 \\
\cite{our_chehade2025iscc} & Sess. & \cellcolor[rgb]{0.47,0.77,0.40}96.59 & \cellcolor[rgb]{0.56,0.81,0.41}96.54 & - & \cellcolor[rgb]{0.00,0.42,0.22}0.09 & \cellcolor[rgb]{0.73,0.88,0.46}20.12 & \cellcolor[rgb]{0.57,0.81,0.41}10.08 \\
\cite{song2019encrypted} & Sess. & - & \cellcolor[rgb]{0.99,0.76,0.45}91.80 & - & \cellcolor[rgb]{0.03,0.46,0.24}0.22 & \cellcolor[rgb]{0.99,0.74,0.43}313.60 & \cellcolor[rgb]{0.99,0.75,0.44}267.22 \\
\cite{centime_maonan2021centime} & Sess. & \cellcolor[rgb]{0.00,0.41,0.22}99.79 & \cellcolor[rgb]{0.00,0.41,0.22}99.80 & - & \cellcolor[rgb]{0.45,0.76,0.39}2.06 & \cellcolor[rgb]{1.00,0.85,0.52}200.70 & \cellcolor[rgb]{0.69,0.04,0.15}3152.27 \\
\cite{zou2024novel_residual} & Sess. & \cellcolor[rgb]{0.01,0.42,0.22}99.64 & \cellcolor[rgb]{0.01,0.43,0.23}99.64 & \cellcolor[rgb]{0.00,0.41,0.22}98.78 & \cellcolor[rgb]{0.56,0.81,0.41}2.82 & \cellcolor[rgb]{0.77,0.90,0.49}25.09 & \cellcolor[rgb]{0.98,0.99,0.72}62.46 \\
\cite{wang2017end} & Sess. & \cellcolor[rgb]{0.65,0.00,0.15}86.60 & - & - & \cellcolor[rgb]{0.82,0.92,0.52}5.83 & \cellcolor[rgb]{0.77,0.90,0.49}25.09 & \cellcolor[rgb]{0.90,0.96,0.61}39.73 \\
\cite{wang2020encrypted} & Sess. & \cellcolor[rgb]{0.18,0.63,0.33}98.00 & \cellcolor[rgb]{0.22,0.66,0.34}98.00 & - & \cellcolor[rgb]{0.84,0.93,0.53}6.17 & \cellcolor[rgb]{0.77,0.90,0.49}25.09 & \cellcolor[rgb]{0.90,0.96,0.61}40.39 \\
\cite{attention_spatio_hu2023network} & Sess. & - & \cellcolor[rgb]{0.20,0.64,0.34}98.10 & - & \cellcolor[rgb]{0.87,0.94,0.57}6.80 & \cellcolor[rgb]{0.77,0.90,0.49}25.09 & \cellcolor[rgb]{0.78,0.91,0.50}23.30 \\
\cite{li2025trustworthy} & Sess. & \cellcolor[rgb]{0.73,0.08,0.15}87.15 & \cellcolor[rgb]{0.65,0.00,0.15}87.93 & - & \cellcolor[rgb]{0.88,0.95,0.59}7.13 & \cellcolor[rgb]{0.92,0.97,0.64}50.18 & \cellcolor[rgb]{0.92,0.96,0.63}43.69 \\
\cite{li2024l2} & Sess. & \cellcolor[rgb]{0.35,0.72,0.38}97.18 & \cellcolor[rgb]{0.29,0.69,0.36}97.73 & - & \cellcolor[rgb]{1.00,0.89,0.56}18.11 & \cellcolor[rgb]{0.94,0.39,0.25}1003.52 & \cellcolor[rgb]{0.65,0.00,0.15}3901.43 \\
\cite{lu2021iclstm} & Sess. & \cellcolor[rgb]{0.15,0.62,0.33}98.10 & \cellcolor[rgb]{0.66,0.86,0.42}96.00 & \cellcolor[rgb]{0.76,0.90,0.49}97.60 & \cellcolor[rgb]{0.98,0.57,0.33}39.12 & \cellcolor[rgb]{1.00,0.90,0.59}151.51 & \cellcolor[rgb]{0.99,0.76,0.45}251.06 \\
\hline
\hline
\cite{yao2019identification} & Flows & \cellcolor[rgb]{0.99,0.81,0.49}91.20 & - & - & \cellcolor[rgb]{0.29,0.69,0.36}1.35 & \cellcolor[rgb]{0.10,0.59,0.31}2.00 & \cellcolor[rgb]{0.04,0.49,0.26}0.86 \\
\cite{xu2024cascaded} & Flows & \cellcolor[rgb]{0.99,0.75,0.44}90.82 & \cellcolor[rgb]{0.99,0.76,0.45}91.83 & - & \cellcolor[rgb]{0.73,0.88,0.46}4.49 & - & - \\
\cite{only_header_cui2022only} & Flows & \cellcolor[rgb]{0.04,0.48,0.25}99.20 & \cellcolor[rgb]{0.05,0.50,0.26}99.20 & - & \cellcolor[rgb]{0.81,0.92,0.52}5.78 & \cellcolor[rgb]{1.00,0.96,0.68}102.40 & \cellcolor[rgb]{0.99,0.64,0.36}400.43 \\
\cite{li2025trustworthy} & Flows & \cellcolor[rgb]{0.77,0.12,0.15}87.39 & \cellcolor[rgb]{0.93,0.38,0.24}89.99 & - & \cellcolor[rgb]{0.88,0.95,0.59}7.13 & \cellcolor[rgb]{0.92,0.97,0.64}50.18 & \cellcolor[rgb]{0.92,0.96,0.63}43.69 \\
\cite{yao2019identification} & Flows & \cellcolor[rgb]{0.97,0.51,0.30}89.50 & - & - & \cellcolor[rgb]{0.97,0.56,0.32}39.49 & \cellcolor[rgb]{0.14,0.62,0.32}2.56 & \cellcolor[rgb]{0.03,0.47,0.25}0.67 \\
\cite{nas_etc_malekghaini2023automl4etc} & Pack. & \cellcolor[rgb]{0.84,0.93,0.54}94.35 & \cellcolor[rgb]{0.91,0.96,0.63}94.40 & - & \cellcolor[rgb]{0.03,0.47,0.25}0.23 & - & - \\
\cite{lotfollahi2020deep} & Pack. & - & \cellcolor[rgb]{0.99,0.79,0.47}92.00 & - & \cellcolor[rgb]{0.13,0.61,0.32}0.81 & \cellcolor[rgb]{0.00,0.41,0.22}0.40 & \cellcolor[rgb]{0.09,0.57,0.30}1.62 \\
\cite{lotfollahi2020deep} & Pack. & - & \cellcolor[rgb]{1.00,0.93,0.63}93.00 & - & \cellcolor[rgb]{0.65,0.85,0.42}3.55 & \cellcolor[rgb]{1.00,0.96,0.69}100.00 & \cellcolor[rgb]{0.99,0.81,0.49}208.19 \\
\cite{cui2019session} & Hybrid & \cellcolor[rgb]{0.05,0.50,0.26}99.10 & \cellcolor[rgb]{0.04,0.48,0.25}99.30 & - & \cellcolor[rgb]{0.87,0.94,0.57}6.80 & \cellcolor[rgb]{1.00,0.96,0.68}102.40 & \cellcolor[rgb]{0.99,0.64,0.36}401.89 \\
\cite{seydali2023cbs} & Hybrid & \cellcolor[rgb]{0.00,0.42,0.22}99.70 & \cellcolor[rgb]{0.04,0.48,0.25}99.30 & - & \cellcolor[rgb]{0.65,0.00,0.15}138.45 & \cellcolor[rgb]{0.65,0.00,0.15}4500.00 & \cellcolor[rgb]{1.00,0.88,0.55}153.87 \\
\hline
\multicolumn{8}{c}{\textbf{USTC2016 Dataset}} \\
\hline
\textbf{\textbf{RepTC-USTC}} & \textbf{\textbf{sess}.} & \cellcolor[rgb]{0.45,0.76,0.39}\textbf{\makecell{99.62\\ $\pm$\textbf{0.02}} } & \cellcolor[rgb]{0.51,0.79,0.40}\textbf{\makecell{99.62\\ $\pm$\textbf{0.02}} } & \cellcolor[rgb]{0.01,0.42,0.22}\textbf{\makecell{99.42\\ $\pm$\textbf{0.03}} } & \cellcolor[rgb]{0.00,0.41,0.22}\textbf{0.02} & \cellcolor[rgb]{0.00,0.41,0.22}\textbf{1.52} & \cellcolor[rgb]{0.00,0.41,0.22}\textbf{0.48} \\
\cite{our_chehade2025tnsm} & Sess. & \cellcolor[rgb]{0.26,0.67,0.35}99.73 & \cellcolor[rgb]{0.32,0.70,0.37}99.73 & - & \cellcolor[rgb]{0.01,0.42,0.22}0.09 & \cellcolor[rgb]{0.65,0.85,0.42}20.12 & \cellcolor[rgb]{0.54,0.80,0.40}10.08 \\
\cite{wang2017malware} & Sess. & \cellcolor[rgb]{0.99,0.99,0.73}99.17 & - & - & \cellcolor[rgb]{0.49,0.78,0.40}3.28 & \cellcolor[rgb]{0.72,0.88,0.46}25.09 & \cellcolor[rgb]{0.80,0.92,0.51}27.86 \\
\cite{li2025trustworthy} & Sess. & \cellcolor[rgb]{1.00,0.92,0.61}99.04 & \cellcolor[rgb]{1.00,0.86,0.53}99.07 & - & \cellcolor[rgb]{0.73,0.89,0.47}7.13 & \cellcolor[rgb]{0.89,0.95,0.60}50.18 & \cellcolor[rgb]{0.90,0.96,0.61}43.69 \\
\cite{li2024l2} & Sess. & \cellcolor[rgb]{0.09,0.57,0.30}99.84 & \cellcolor[rgb]{0.15,0.62,0.33}99.81 & - & \cellcolor[rgb]{0.97,0.99,0.71}18.11 & \cellcolor[rgb]{0.89,0.30,0.20}1003.52 & \cellcolor[rgb]{0.65,0.00,0.15}3901.43 \\
\hline
\hline
\cite{nas_zhang2023automatic} & Flows & \cellcolor[rgb]{0.00,0.41,0.22}99.99 & \cellcolor[rgb]{0.00,0.41,0.22}99.99 & - & \cellcolor[rgb]{0.01,0.42,0.22}0.07 & - & \cellcolor[rgb]{0.72,0.88,0.46}19.63 \\
\cite{xu2024cascaded} & Flows & \cellcolor[rgb]{0.65,0.00,0.15}98.32 & \cellcolor[rgb]{0.65,0.00,0.15}98.49 & - & \cellcolor[rgb]{0.59,0.82,0.41}4.49 & - & - \\
\cite{li2025trustworthy} & Flows & \cellcolor[rgb]{1.00,0.96,0.69}99.10 & \cellcolor[rgb]{1.00,0.93,0.63}99.15 & - & \cellcolor[rgb]{0.73,0.89,0.47}7.13 & \cellcolor[rgb]{0.89,0.95,0.60}50.18 & \cellcolor[rgb]{0.90,0.96,0.61}43.69 \\
\cite{yao2019capsule} & Flows & \cellcolor[rgb]{0.68,0.03,0.15}98.35 & - & - & \cellcolor[rgb]{0.65,0.00,0.15}450.58 & \cellcolor[rgb]{0.65,0.00,0.15}2816.00 & \cellcolor[rgb]{0.88,0.26,0.19}1402.37 \\
\cite{ieeeaccess_wang2020deep} & Pack. & \cellcolor[rgb]{0.03,0.46,0.24}99.94 & - & - & \cellcolor[rgb]{0.03,0.46,0.24}0.22 & \cellcolor[rgb]{0.53,0.80,0.40}13.82 & \cellcolor[rgb]{0.18,0.63,0.33}3.06 \\
\cite{yu2025model} & Pack. & \cellcolor[rgb]{0.02,0.45,0.24}99.95 & \cellcolor[rgb]{0.61,0.83,0.41}99.56 & - & \cellcolor[rgb]{0.34,0.71,0.37}2.10 & \cellcolor[rgb]{0.25,0.67,0.35}6.27 & \cellcolor[rgb]{0.75,0.89,0.48}22.60 \\
\hline
\multicolumn{8}{c}{\textbf{Edge-IIoTset Dataset}} \\
\hline
\textbf{\textbf{RepTC-Edge}} & \textbf{\textbf{sess}.} & \cellcolor[rgb]{0.00,0.41,0.22}\textbf{\makecell{99.97\\ $\pm$\textbf{0.02}} } & \cellcolor[rgb]{0.02,0.45,0.24}\textbf{\makecell{99.97\\ $\pm$\textbf{0.02}} } & \cellcolor[rgb]{0.02,0.45,0.24}\textbf{\makecell{98.29\\ $\pm$\textbf{0.18}} } & \cellcolor[rgb]{0.00,0.41,0.22}\textbf{0.02} & \cellcolor[rgb]{0.06,0.53,0.28}\textbf{1.14} & \cellcolor[rgb]{0.06,0.52,0.27}\textbf{0.26} \\
\hline
\hline
\cite{nas_zhang2023automatic} & Flows & \cellcolor[rgb]{0.01,0.42,0.22}99.94 & \cellcolor[rgb]{0.65,0.00,0.15}98.84 & - & \cellcolor[rgb]{0.03,0.47,0.25}0.07 & - & \cellcolor[rgb]{0.65,0.00,0.15}19.67 \\
\cite{alshehri2024self} & Features & \cellcolor[rgb]{0.00,0.42,0.22}99.95 & \cellcolor[rgb]{0.86,0.94,0.55}99.53 & - & \cellcolor[rgb]{0.03,0.47,0.25}0.07 & \cellcolor[rgb]{0.46,0.77,0.39}1.86 & \cellcolor[rgb]{0.63,0.84,0.41}1.50 \\
\cite{de2022hybrid} & Features & \cellcolor[rgb]{0.65,0.00,0.15}97.14 & - & \cellcolor[rgb]{0.65,0.00,0.15}74.62 & \cellcolor[rgb]{0.07,0.53,0.28}0.12 & \cellcolor[rgb]{0.00,0.41,0.22}0.90 & \cellcolor[rgb]{0.00,0.41,0.22}0.05 \\
\cite{ullah2024attention} & Features & \cellcolor[rgb]{0.00,0.42,0.22}99.95 & \cellcolor[rgb]{0.73,0.89,0.47}99.60 & - & \cellcolor[rgb]{0.28,0.68,0.36}0.27 & \cellcolor[rgb]{0.95,0.98,0.68}3.46 & \cellcolor[rgb]{0.92,0.97,0.64}2.97 \\
\cite{hassini2024end} & Features & \cellcolor[rgb]{0.00,0.41,0.22}99.96 & \cellcolor[rgb]{0.00,0.41,0.22}100.00 & \cellcolor[rgb]{0.00,0.41,0.22}99.00 & \cellcolor[rgb]{0.36,0.72,0.38}0.32 & \cellcolor[rgb]{0.89,0.30,0.20}8.06 & \cellcolor[rgb]{1.00,0.84,0.51}5.65 \\
\cite{selem2025deep} & Features & \cellcolor[rgb]{1.00,0.83,0.51}98.20 & - & - & \cellcolor[rgb]{0.65,0.00,0.15}2.99 & \cellcolor[rgb]{0.65,0.00,0.15}10.84 & \cellcolor[rgb]{0.97,0.55,0.32}8.84 \\
\hline

    \end{tabular}%
    }
\end{table}


Empirical evidence shows that the models derived by RepTC achieve a favorable accuracy--complexity trade-off relative to larger fixed-design models. On ISCX, an encrypted-traffic dataset often addressed with relatively complex models, \textit{RepTC-ISCX} achieved an average accuracy of 96.21\%, F1\textsubscript{w} of 96.24\%, and F1\textsubscript{m} of 95.33\%, with only 0.05M parameters, 2.55K tensor elements, and 0.30M FLOPs.

Within the session-based literature, higher predictive scores have been reported on ISCX, but typically with much larger architectures. The models in~\cite{zou2024novel_residual} and~\cite{centime_maonan2021centime}, for instance, reached 99.64\% and 99.79\% accuracy, but required about 208$\times$ and 10{,}507$\times$ more FLOPs, respectively. Likewise, the session models in~\cite{lu2021iclstm} and~\cite{wang2020encrypted} achieved 98.10\% and 98.00\% accuracy, while requiring up to 782$\times$ and 123$\times$ more parameters, and 836$\times$ and 134$\times$ more FLOPs, respectively. Results in~\cite{wang2017end,song2019encrypted,li2025trustworthy} also show that such larger models may still yield clearly lower predictive performance, with accuracy or F1\textsubscript{w} remaining below 92\% in some cases. Compared with the HW-NAS baselines in~\cite{our_chehade2025tnsm,our_chehade2025iscc}, which achieved 96.60\% and 96.59\% accuracy, respectively, \textit{RepTC-ISCX} required about 1.8$\times$ fewer parameters, 8$\times$ smaller intermediate tensors, and over 33$\times$ fewer FLOPs. This comparison indicates that optimizing network structure alone may not fully capture the efficiency--performance design space, and that jointly optimizing architecture, input size, and header preprocessing can yield a more favorable trade-off under strict hardware constraints; for \textit{RepTC-ISCX}, this comes with a  moderate accuracy gap relative to the heaviest session-based models but enables substantial reductions in memory and computation that are critical for deployment on sub-watt edge platforms.

Methods based on flows, packets, or hybrid inputs provide additional context on ISCX. The flow-based models in~\cite{yao2019identification} reached 91.2\% and 89.5\% accuracy with 0.86M and 0.67M FLOPs, respectively, whereas~\cite{li2025trustworthy} reported both session- and flow-level variants on this dataset, each with 7.13M parameters and 43.69M FLOPs, with the flow-level version reaching 87.39\% accuracy and remaining roughly 145$\times$ heavier in computation than \textit{RepTC-ISCX}. Packet-level models in~\cite{lotfollahi2020deep} reported F1\textsubscript{w} up to 93.00\%, with configurations ranging from 0.81M parameters and 1.62M FLOPs to 3.55M parameters and 208.19M FLOPs. Hybrid models such as~\cite{seydali2023cbs,cui2019session} reached accuracy up to 99.70\%, but required 153.87M and 401.89M FLOPs, corresponding to more than 512$\times$ and 1339$\times$ the compute cost of \textit{RepTC-ISCX}.

\textit{RepTC-USTC} achieved 99.62\% accuracy, 99.62\% F1\textsubscript{w}, and 99.42\% F1\textsubscript{m}, with 0.02M parameters, 1.52K tensor elements, and 0.48M FLOPs. Average results on USTC are generally around 99\% accuracy, and our proposal remained within that range while preserving very low complexity. In the session-based group, the model in~\cite{li2024l2} reported slightly higher accuracy, but required 18.11M parameters and 3901.43M FLOPs.
Compared with~\cite{our_chehade2025tnsm}, which achieved 99.73\% accuracy with 0.09M parameters and 10.08M FLOPs, \textit{RepTC-USTC} remained notably more compact and efficient. Among non-session methods,~\cite{yu2025model} required 2.10M parameters and 22.60M FLOPs, about 47$\times$ more compute than the proposed model, while~\cite{yao2019capsule} and~\cite{ieeeaccess_wang2020deep} also had higher complexity. Overall, \textit{RepTC-USTC} was the most efficient model in the comparison while retaining highly competitive predictive performance.

Comparable outcomes were observed on Edge-IIoTset: \textit{RepTC-Edge} reached 99.97\% accuracy, 99.97\% F1\textsubscript{w}, and 98.29\% F1\textsubscript{m}, showing negligible variation across runs, with just 0.02M parameters and 0.26M FLOPs. The high F1\textsubscript{m} also indicates strong performance across classes despite dataset imbalance.  Several competing methods on this dataset also reported performance in the 99.9\% range, generally with higher architectural cost. By comparison, the method in \cite{selem2025deep} required 2.99M parameters and 8.84M FLOPs (over 34$\times$ more compute); \cite{hassini2024end} used 0.32M parameters and 5.65M FLOPs (22$\times$); and \cite{ullah2024attention} reached 2.97M FLOPs and 3.46K tensor elements, corresponding to an 11$\times$ increase in compute cost. The model in \cite{alshehri2024self} was more compact than the other baselines
(0.07M parameters, 1.50M FLOPs), but still required about
\(2.8\times\) more parameters and \(5.8\times\) more FLOPs
than RepTC-Edge. Finally,~\cite{de2022hybrid} required fewer FLOPs (0.05M) and a smaller
peak tensor (0.90K), but used about \(5\times\) more parameters and achieved
lower F1\textsubscript{m} (74.62\%) and accuracy (97.14\%).
These methods also relied on statistical traffic features, implying
explicit feature engineering at inference time.

The research presented in~\cite{nas_zhang2023automatic} covered NAS-searched flow-based models and yielded remarkable accuracy (99.99\% on USTC, 99.94\% on Edge-IIoTset), comparable to those scored by the \textit{RepTC} variants (99.62\% and 99.97\%, respectively). At the same time, the associated models required about 19.6M FLOPs, thus proving roughly 40$\times$ and 75$\times$ heavier than \textit{RepTC-USTC} and \textit{RepTC-Edge}, respectively. Similarly,~\cite{nas_etc_malekghaini2023automl4etc} applied NAS to packet-level ISCX, giving 94.35\% accuracy with 0.23M parameters; FLOPs and tensor usage were not reported. \textit{RepTC-ISCX} achieved 96.21\% with 0.05M parameters and 0.30M FLOPs.
RepTC adopts a representation-aware design perspective, jointly optimizing model architecture, input size, and header preprocessing within a unified search space. This joint formulation enables improved efficiency compared with approaches that optimize network structure alone, while maintaining competitive predictive performance.

\subsection{Ablation Study}

\subsubsection{\deleted{Constrained Search Analysis}\added{Joint Versus Decoupled Search Analysis}
}

We ran the RepTC search under multiple constrained configurations
to isolate the effect of each optimized component
\added{and to compare joint optimization with a sequential alternative}.
\deleted{Four search configurations were evaluated: i) architecture search only;
ii) architecture + header preprocessing; iii) architecture + input length; and
iv) the full RepTC search over architecture, header preprocessing, and input length.}
\added{Five search configurations were evaluated: i) architecture search only;
ii) architecture + header preprocessing; iii) architecture + input length;
iv) a two-stage decoupled representation-to-architecture
(\(R\!\rightarrow\!a\)) search, where \(R=(S_h,L_s)\), which first searches
the representation with the architecture fixed, then fixes that representation
and searches the architecture; and v) the full RepTC joint search over
architecture, header preprocessing, and input length.}

This analysis was performed on the ISCX VPN-nonVPN dataset, a widely used ETC
benchmark supporting a controlled comparison across search configurations.
When a component is fixed
\added{in the restricted search configurations},
\(S_h\) is set to \(S_2\) from Table~\ref{tab:preprocessing_strategies} and
\(L_s\) to 784\,B, in line with the common practice summarized in
Table~\ref{tab:tc_representation_comparison}, where Ethernet removal with
IP anonymization and a 784-byte input are among the most common
fixed settings in prior raw-traffic TC studies.

\added{For the decoupled \(R\!\rightarrow\!a\) baseline, the first stage starts from \(a_0\), the architecture selected by the architecture-only search, keeps it fixed, and searches \(R=(S_h,L_s)\). The second stage then fixes the representation selected in the first stage and searches \(a\). This provides a direct sequential counterpart to RepTC's joint optimization.}

All search configurations used the same search budget, training protocol,
and hardware constraints described in Section~\ref{setup} \superadded{and were
executed on the same computing platform}.
\added{For the decoupled \(R\!\rightarrow\!a\) baseline, each stage independently
used this same full search budget, so that neither representation nor
architecture optimization was restricted by a smaller allocation.}
The best candidate returned by each search was then retrained and evaluated
under identical conditions. Table~\ref{tab:constrained_search} summarizes the
resulting predictive and architectural metrics
\added{together with the corresponding wall-clock time,} \superadded{measured as the total elapsed search time};
predictive results are reported as mean \(\pm\) standard deviation over
10 independent runs.
A checkmark (\cmark) indicates that the corresponding component is optimized
during the search, while fixed components are reported explicitly.
\added{For the decoupled \(R\!\rightarrow\!a\) baseline, the two rows report
its first-stage representation search (\(R\)) and second-stage architecture
search (\(a\)), respectively.}
Acc. denotes accuracy; F1\textsubscript{m} and F1\textsubscript{w} denote
macro- and weighted-F1 scores, respectively; Par. indicates the number of
parameters; M-Tens. the maximum intermediate tensor size;
\added{and Wall-clock reports the total search time.}

\begin{table}[htbp]
\centering
\caption{
\added{Search ablation results.}
}
\label{tab:constrained_search}
\renewcommand{\arraystretch}{1.0}
\large
\resizebox{\columnwidth}{!}{%
\begin{tabular}{
>{\centering\arraybackslash}c|
>{\centering\arraybackslash}c|
>{\centering\arraybackslash}c|
>{\centering\arraybackslash}c||
>{\centering\arraybackslash}c|
>{\centering\arraybackslash}c|
>{\centering\arraybackslash}c|
>{\centering\arraybackslash}c|
>{\centering\arraybackslash}c|
>{\centering\arraybackslash}c||
>{\centering\arraybackslash}c
}
\hline

\multirow{2}{*}{\textbf{Setting}} &
\multicolumn{3}{c||}{\textbf{Search Over}} &
\multicolumn{6}{c||}{\textbf{Results}} &
\multirow{2}{*}{\makecell{\textbf{Wall-}\\\textbf{clock (h)}}} \\
\cline{2-10}

& \textbf{\(a\)}
& \textbf{\(S_h\)}
& \textbf{\(L_s\)}
& \makecell{\textbf{Acc.}\\(\%)}
& \makecell{\textbf{F1\textsubscript{m}}\\(\%)}
& \makecell{\textbf{F1\textsubscript{w}}\\(\%)}
& \makecell{\textbf{Par.}\\(M)}
& \makecell{\textbf{M-Tens.}\\(K)}
& \makecell{\textbf{FLOPs}\\(M)}
& \\
\hline

\multicolumn{11}{c}{\added{\textit{Restricted search ablations}}} \\
\hline

\makecell{Architecture\\search}
& \cmark
& \(S_2\)
& 784\,B
& \makecell{92.74\\\small$\pm$0.76}
& \makecell{92.33\\\small$\pm$0.58}
& \makecell{92.79\\\small$\pm$0.74}
& 0.03
& 3.02
& 0.50
& \added{17.76} \\
\hline

\makecell{Architecture+\\\(S_h\) search}
& \cmark
& \cmark
& 784\,B
& \makecell{93.12\\\small$\pm$0.77}
& \makecell{92.45\\\small$\pm$0.67}
& \makecell{93.15\\\small$\pm$0.75}
& 0.03
& 2.46
& 0.26
& \added{16.76} \\
\hline

\makecell{Architecture+\\\(L_s\) search}
& \cmark
& \(S_2\)
& \cmark
& \makecell{93.39\\\small$\pm$0.80}
& \makecell{92.69\\\small$\pm$0.74}
& \makecell{93.44\\\small$\pm$0.78}
& 0.04
& 1.12
& 0.46
& \added{17.26} \\
\hline

\multicolumn{11}{c}{\added{\textit{Decoupled \(R\!\rightarrow\!a\) search}}} \\
\hline

\added{\makecell{Stage 1:\\\(R\) search}}
& \added{\(a_0\)}
& \added{\cmark}
& \added{\cmark}
& \added{\makecell{93.23\\\small$\pm$0.61}}
& \added{\makecell{93.15\\\small$\pm$0.86}}
& \added{\makecell{92.77\\\small$\pm$0.31}}
& \added{0.03}
& \added{2.27}
& \added{0.32}
& \added{16.50} \\
\hline

\added{\makecell{Stage 2:\\\(a\) search}}
& \added{\cmark}
& \added{\(S_1\)}
& \added{328\,B}
& \added{\makecell{94.16\\\small$\pm$0.25}}
& \added{\makecell{94.12\\\small$\pm$0.40}}
& \added{\makecell{93.05\\\small$\pm$0.36}}
& \added{0.05}
& \added{2.69}
& \added{0.46}
& \added{16.22} \\
\hline

\multicolumn{11}{c}{\added{\textit{Joint search}}} \\
\hline

\textbf{RepTC}
& \cmark
& \cmark
& \cmark
& \makecell{\textbf{96.21}\\\small\textbf{$\pm$0.51}}
& \makecell{\textbf{95.33}\\\small\textbf{$\pm$0.48}}
& \makecell{\textbf{96.24}\\\small\textbf{$\pm$0.51}}
& \textbf{0.05}
& \textbf{2.55}
& \textbf{0.30}
& \added{\textbf{18.18}} \\
\hline

\end{tabular}%
}
\end{table}

The constrained searches selected \(S_h=S_1\) for the architecture+\(S_h\)
configuration and \(L_s=358\) for the architecture+\(L_s\) configuration.
\added{For the decoupled \(R\!\rightarrow\!a\) baseline, the first stage
selected \(S_h=S_1\) and \(L_s=328\,\mathrm{B}\), which were then fixed
during the second-stage architecture search.}
\deleted{In all cases, the resulting models remained inferior to the full RepTC
solution.}\added{The restricted and sequential alternatives yielded weaker
accuracy--resource trade-offs than the full joint RepTC search under the
evaluated search setting.}

\added{The wall-clock times were comparable across the single-stage searches,
ranging from 16.76 to 18.18\,h, with RepTC requiring 18.18\,h compared with
17.76\,h for architecture-only search, corresponding to approximately
\(2.4\%\) additional wall-clock time. The two decoupled stages required
16.50\,h and 16.22\,h; since each stage was intentionally given the same full
search budget, the complete sequential procedure required 32.72\,h
(\superadded{50.48\,h if the preceding architecture-only search used to obtain
\(a_0\) is also included}).}
\added{The second decoupled stage improves upon the representation-search stage, but the final sequential model remains below the jointly optimized RepTC solution.}
These results indicate that configurations achieving this balance would not emerge without jointly optimizing architecture, preprocessing strategy, and input length.

\begin{figure}[htbp]
    \centering
    \includegraphics[width=\columnwidth]
    {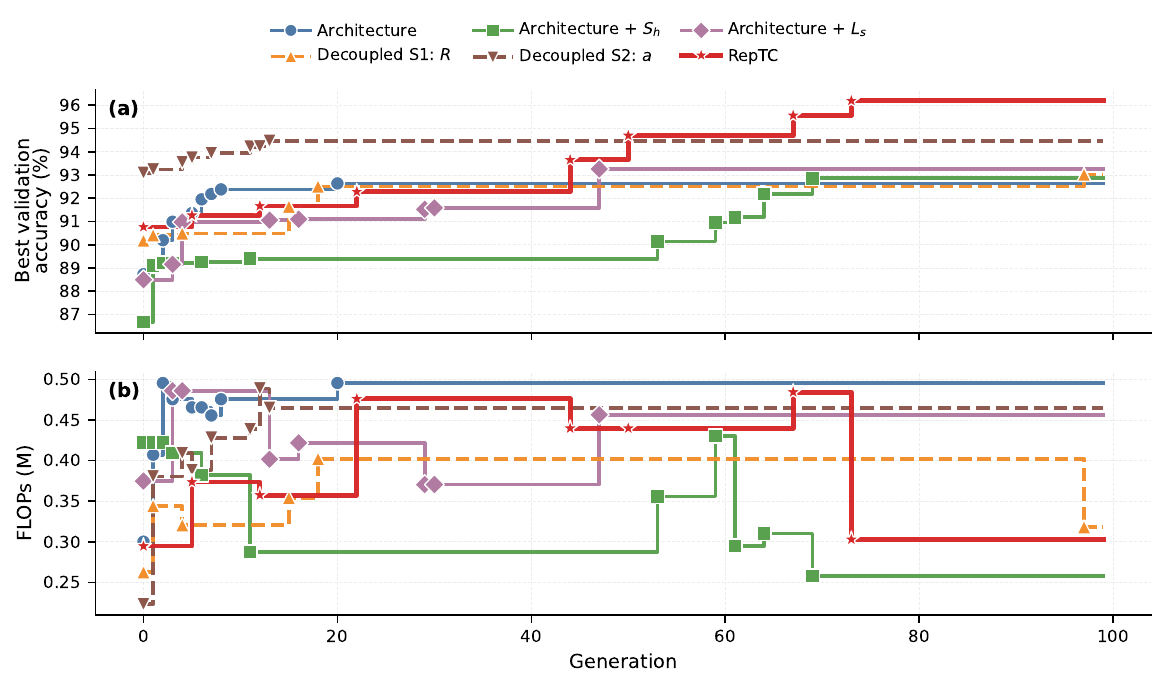}
    \caption{\added{Convergence of the evaluated search configurations.}}
    \label{fig:search_convergence}
\end{figure}

\added{We further assessed the search configurations in
Table~\ref{tab:constrained_search} using the convergence profiles in
Fig.~\ref{fig:search_convergence}. Figure~\ref{fig:search_convergence}(a) shows the best validation accuracy
found up to each generation, while Fig.~\ref{fig:search_convergence}(b) shows
the FLOPs of the corresponding best-so-far candidate\superadded{; the FLOPs change when a new best candidate has a
different computational cost}. The two decoupled
stages are shown separately; Stage~2 starts from the representation selected at
the end of Stage~1 and then searches the architecture while keeping that
representation fixed. Symbols mark generations at which a new best solution is
found.}

\superadded{The convergence profiles show different search dynamics across the
evaluated configurations. Architecture-only search stabilizes earlier, whereas
RepTC continues improving later in the search and attains the highest validation
accuracy. The decoupled stages follow different trajectories but remain below
the final RepTC solution.}

\subsubsection{\added{Resource-Constraint Sensitivity}}

\added{To further examine how the hardware budget affects the solutions selected
by RepTC, we performed a one-factor-at-a-time sensitivity analysis on the ISCX
VPN-nonVPN dataset to assess how RepTC adapts around the reference deployment
setting. We varied three resource constraints independently: maximum
intermediate tensor size (M-Tens.), which determines peak RAM demand; parameter
count (Par.), which directly affects Flash storage; and FLOPs, which controls
computational cost. For each experiment, only the constraint under study was
changed, while the remaining search settings and hardware thresholds were kept
identical to those used for RepTC.}

\added{For each resource, we considered constrained and relaxed budgets
relative to the reference RepTC setting. The constrained threshold was halved,
while the relaxed threshold was doubled. Thus, M-Tens. varied from 4K to 2K
and 8K, FLOPs from 0.50M to 0.25M and 1.00M, and parameters from 0.06M to
0.03M and 0.12M. \added{The relaxed limits remain within the resource envelope
of the target hardware considered in Section~\ref{deployment}.}
Table~\ref{tab:resource_sensitivity} reports the selected representation and its
predictive and hardware characteristics under each budget.}

\added{The results show a clear effect of the active hardware constraint on the
solution selected by RepTC. Tightening the M-Tens., FLOPs, or parameter limit
leads the search to a model that uses less of the corresponding resource, with
a reduction in predictive performance. Relaxing the resource limits generally
improves performance relative to the constrained settings, although increasing
an individual budget does not necessarily improve upon the reference RepTC
configuration.}

\added{The selected representation also responds to the resource budget.
Changes in the M-Tens. and FLOPs limits lead to different choices of \(S_h\),
while \(L_s\) varies across the evaluated settings. Despite these changes, the
search repeatedly favors a small subset of preprocessing strategies and, in
most cases, relatively short input lengths. This suggests
that the representation adapts to the available resources without changing
arbitrarily across configurations.}

\begin{table}[htbp]
\centering
\caption{\added{Sensitivity to individual resource constraints.}}
\label{tab:resource_sensitivity}
\renewcommand{\arraystretch}{1.0}
\large
\resizebox{\columnwidth}{!}{%
\begin{tabular}{
>{\centering\arraybackslash}c|
>{\centering\arraybackslash}c||
>{\centering\arraybackslash}c|
>{\centering\arraybackslash}c||
>{\centering\arraybackslash}c|
>{\centering\arraybackslash}c|
>{\centering\arraybackslash}c||
>{\centering\arraybackslash}c|
>{\centering\arraybackslash}c|
>{\centering\arraybackslash}c
}
\hline
\textbf{Level} &
\textbf{Limit} &
\textbf{\(S_h\)} &
\textbf{\(L_s\)} &
\makecell{\textbf{Acc.}\\(\%)} &
\makecell{\textbf{F1\textsubscript{m}}\\(\%)} &
\makecell{\textbf{F1\textsubscript{w}}\\(\%)} &
\makecell{\textbf{Par.}\\(M)} &
\makecell{\textbf{M-Tens.}\\(K)} &
\makecell{\textbf{FLOPs}\\(M)} \\
\hline

\multicolumn{10}{c}{\added{\textit{Maximum tensor size / RAM}}} \\
\hline

\added{Constrained}
& \added{2K}
& \added{\(S_4\)}
& \added{328\,B}
& \added{\makecell{93.65\\\small$\pm$0.21}}
& \added{\makecell{92.55\\\small$\pm$0.45}}
& \added{\makecell{93.44\\\small$\pm$0.25}}
& \added{0.04}
& \added{1.27}
& \added{0.26} \\
\hline

\added{Relaxed}
& \added{8K}
& \added{\(S_2\)}
& \added{742\,B}
& \added{\makecell{96.12\\\small$\pm$0.32}}
& \added{\makecell{95.54\\\small$\pm$0.65}}
& \added{\makecell{96.10\\\small$\pm$0.27}}
& \added{0.04}
& \added{4.27}
& \added{0.46} \\
\hline

\multicolumn{10}{c}{\added{\textit{Computational cost / FLOPs}}} \\
\hline

\added{Constrained}
& \added{0.25M}
& \added{\(S_4\)}
& \added{234\,B}
& \added{\makecell{92.65\\\small$\pm$0.42}}
& \added{\makecell{92.55\\\small$\pm$0.61}}
& \added{\makecell{92.44\\\small$\pm$0.33}}
& \added{0.05}
& \added{0.91}
& \added{0.24} \\
\hline

\added{Relaxed}
& \added{1.00M}
& \added{\(S_1\)}
& \added{236\,B}
& \added{\makecell{96.65\\\small$\pm$0.20}}
& \added{\makecell{96.05\\\small$\pm$0.55}}
& \added{\makecell{96.54\\\small$\pm$0.17}}
& \added{0.06}
& \added{2.80}
& \added{0.76} \\
\hline

\multicolumn{10}{c}{\added{\textit{Parameter count / Flash}}} \\
\hline

\added{Constrained}
& \added{0.03M}
& \added{\(S_1\)}
& \added{332\,B}
& \added{\makecell{91.65\\\small$\pm$0.29}}
& \added{\makecell{90.55\\\small$\pm$0.57}}
& \added{\makecell{91.44\\\small$\pm$0.34}}
& \added{0.02}
& \added{3.80}
& \added{0.32} \\
\hline

\added{Relaxed}
& \added{0.12M}
& \added{\(S_1\)}
& \added{294\,B}
& \added{\makecell{95.65\\\small$\pm$0.16}}
& \added{\makecell{94.85\\\small$\pm$0.45}}
& \added{\makecell{95.44\\\small$\pm$0.14}}
& \added{0.06}
& \added{2.80}
& \added{0.28} \\
\hline

\end{tabular}%
}
\end{table}

\subsubsection{Sensitivity Analysis of Input Representation}
\label{input_sensitivity}

\begin{figure}[h]
    \centering
    \includegraphics[width=\columnwidth]{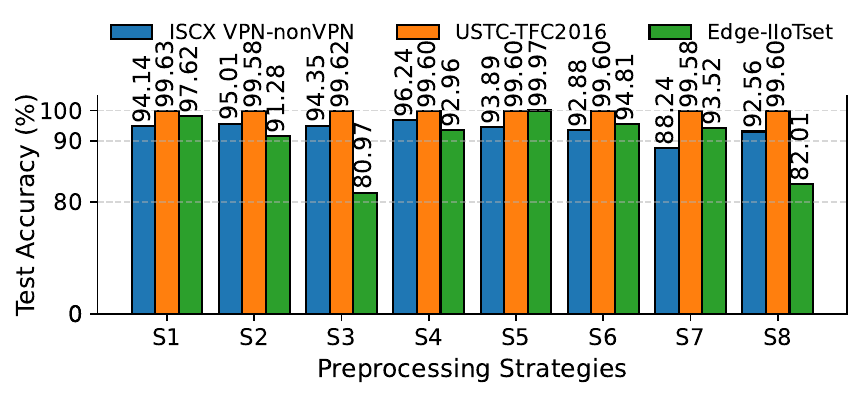}
\caption{Accuracy impact of preprocessing strategy per dataset.}

    \label{fig:ablation_strategies}
\end{figure}

This section analyzes the effect of varying the preprocessing strategy \(S_h\)
and the input length \(L_s\) on the final RepTC models, with all other search
parameters kept fixed.
\added{Besides quantifying representation sensitivity, varying \(S_h\) also
provides a controlled assessment of the extent to which classification depends
on volatile endpoint information such as IP addresses and transport ports.}
For each preprocessing strategy \(S_h\), the optimized RepTC architecture was
kept fixed, while its weights were retrained from scratch using the
corresponding preprocessed training data.
Figure~\ref{fig:ablation_strategies} shows the resulting test accuracy across
the eight strategies defined in Table~\ref{tab:preprocessing_strategies}.

\textit{RepTC-USTC} showed almost constant accuracy across strategies, indicating limited reliance on header fields.
\textit{RepTC-ISCX}, instead, proved sensitive to header processing: strategies removing the Ethernet layer and zeroing IP fields (e.g., \(S_4\)) consistently outperformed those retaining Ethernet with anonymized addresses (e.g., \(S_5\)--\(S_6\)), indicating that generalization improved when low-level link information was discarded. 
On \textit{RepTC-Edge}, performance varied markedly across strategies, with \(S_5\) (Ethernet kept, MAC and port fields zeroed) producing the best results.
\deleted{This suggests that, in industrial traffic, keeping structural headers while masking address fields helps capture protocol-level distinctions.}
\added{Classification also remains feasible under the identifier-stripped
settings \(S_3\) and \(S_7\); \(S_3\) removes the Ethernet header and zeros
IP addresses and transport ports, whereas \(S_7\) zeros MAC addresses, IP
addresses, and transport ports. \superadded{\(S_3\) remains close to the
best-performing configuration on ISCX and USTC, while the fully
identifier-stripped \(S_7\) setting still achieves 93.52\% accuracy on
Edge-IIoTset,} showing that useful discriminative information remains after
removing the considered volatile endpoint identifiers.}

\begin{figure}[t]
    \centering
    \includegraphics[width=\columnwidth]{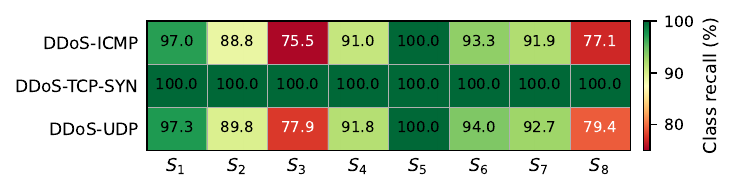}
    \caption{\added{Recall of Edge-IIoTset DDoS classes across \(S_h\).}}
    \label{fig:edge_class_sensitivity}
\end{figure}

\added{To identify the source of the much larger \(S_h\)-induced variation on Edge-IIoTset, we further examined class-level recall across all eight preprocessing strategies. As shown in Fig.~\ref{fig:dataset_distribution}, DDoS-ICMP,
DDoS-TCP-SYN, and DDoS-UDP dominate Edge-IIoTset; together, they also
account for approximately 97.5\% of the held-out test samples. Figure~\ref{fig:edge_class_sensitivity} therefore reports the recall of these three high-support classes across \(S_1\)--\(S_8\).}

\added{The class-level results show that the large \(S_h\)-induced accuracy
variation is concentrated in DDoS-ICMP and DDoS-UDP, whereas DDoS-TCP-SYN
remains essentially invariant. Under \(S_3\), ICMP and UDP recall fall to
75.45\% and 77.90\%, respectively, while TCP-SYN remains at 100\%. Relative to \(S_5\), the ICMP and UDP recall reductions contribute 9.53 and
9.50 percentage points, respectively, accounting for essentially the full
19.00-point accuracy decrease observed under \(S_3\). The same pattern
dominates under the other preprocessing strategies. 
}
\superadded{Thus, the strong \(S_h\) effect on Edge-IIoTset is concentrated
mainly in two high-support attack classes, showing that the degradation is not
uniform across classes.}

\begin{figure}[h]
    \centering
    \includegraphics[width=\columnwidth]{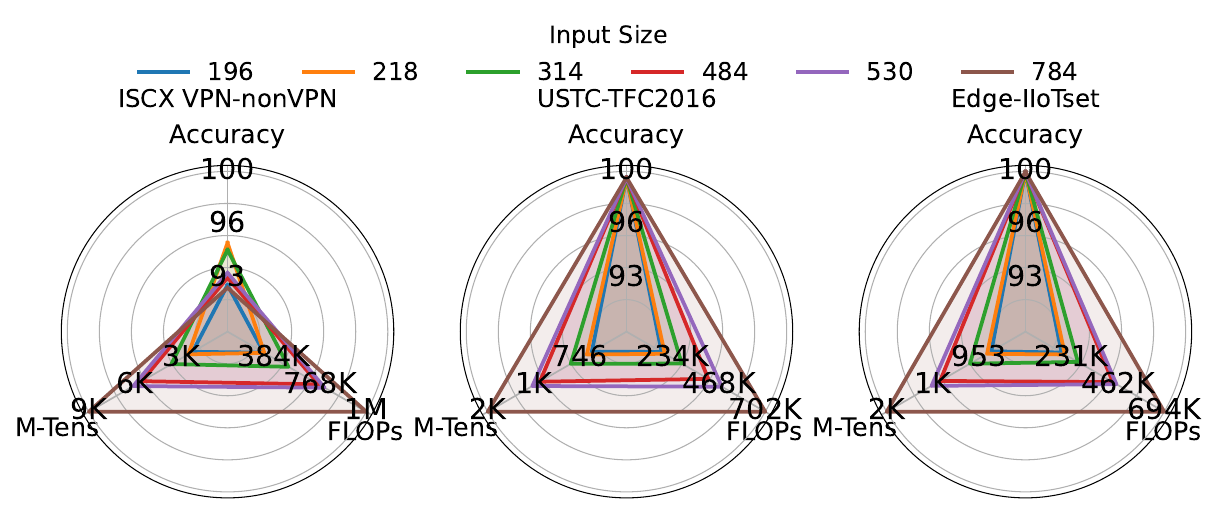}
\caption{Resource–accuracy tradeoffs across input lengths.}
    \label{fig:ablation_input}
\end{figure}

Figure~\ref{fig:ablation_input} illustrates the impact of \( L_s \in [196, 784] \) on accuracy, FLOPs, and memory usage (M-Tens), with \( S_h \) and architecture fixed per respective RepTC model. For each dataset, a radar plot compares performance versus \(L_s\). 

For \textit{RepTC-ISCX}, the highest accuracy occurred when \(L_s\) was 218 bytes, and declined progressively when \(L_s\) increased, despite higher resource usage. In contrast, accuracy remained stable for \textit{RepTC-USTC} and \textit{RepTC-Edge}, and longer inputs did not improve performance. 
Overall, the ablation confirms the dataset-specific patterns identified in the search-space analysis of Section~\ref{sec:statistical}: USTC models were relatively insensitive to \(S_h\) and \(L_s\), ISCX showed combined dependence on both, and Edge-IIoTset remained highly affected by header transformations. This confirms the need to match each architecture with an appropriate input representation, which was the main goal of the joint optimization strategy in RepTC.

\subsubsection{Behavioral Feature Analysis}
\label{sec:behavioral_ablation}

\added{We also examine whether explicit behavioral features provide useful
information beyond the selected raw-byte representation. Specifically, we
consider packet size and inter-arrival time (IAT), while restricting both to
the same observation horizon as RepTC, i.e., to packets required to form the
corresponding \(L_s\)-byte session prefix. Packet size is signed to encode
direction.}

\added{Two behavioral-only settings are considered. IAT-only uses timing
information, whereas Size+IAT combines signed packet size and timing without
raw packet headers or payload bytes. We also evaluate RepTC+IAT
and RepTC+Size+IAT through late fusion, \superadded{where the embedding of the
selected RepTC backbone (obtained after global average pooling) is concatenated
with a 16-unit behavioral embedding and passed through a 32-unit ReLU layer
before classification. The behavioral branch uses two 1-D convolutional layers
with 16 filters and kernel size 3, followed by masked global average pooling
and a 16-unit ReLU layer.} Its architecture is fixed across experiments and
excluded from the RepTC search.}

\added{Table~\ref{tab:behavioral_ablation} reports accuracy and macro-F1 over
10 runs, with standard deviation, parameter count, maximum tensor size
(M-Tens), and FLOPs. The best predictive values within each dataset block are
shown in bold.}


\added{The behavioral-only settings remain substantially below RepTC across all
three datasets. Size+IAT is more informative than IAT alone, with the largest
gain observed on USTC-TFC2016. \superadded{The evaluated late-fusion variants do not improve over RepTC.
Both accuracy and macro-F1 decrease on ISCX VPN-nonVPN and USTC-TFC2016,
while Edge-IIoTset retains similar accuracy with lower macro-F1.}
Explicit behavioral inputs also require additional packet-level feature
extraction and model computation at inference time.}

\begin{table}[htbp]
\centering
\caption{\added{Behavioral-feature ablation and model complexity.}}
\label{tab:behavioral_ablation}

\renewcommand{\arraystretch}{0.82}
\small

\begin{adjustbox}{max width=\columnwidth}
\begin{tabular}{
>{\centering\arraybackslash}c|
>{\centering\arraybackslash}c||
>{\centering\arraybackslash}c|
>{\centering\arraybackslash}c||
>{\centering\arraybackslash}c|
>{\centering\arraybackslash}c|
>{\centering\arraybackslash}c
}
\hline

\textbf{Dataset}
& \textbf{Model}
& \makecell{\textbf{Acc.}\\(\%)}
& \makecell{\textbf{F1\textsubscript{m}}\\(\%)}
& \makecell{\textbf{Par.}\\(M)}
& \makecell{\textbf{M-Tens.}\\(K)}
& \makecell{\textbf{FLOPs}\\(M)}
\\
\hline

\multirow{5}{*}{\makecell{\textbf{ISCX}\\\textbf{VPN-nonVPN}}}

& IAT-only
& \makecell{20.18\\[-1pt]\footnotesize$\pm$0.34}
& \makecell{17.25\\[-1pt]\footnotesize$\pm$0.62}
& 0.001
& 0.064
& 0.008
\\
\cline{2-7}

& Size+IAT
& \makecell{32.95\\[-1pt]\footnotesize$\pm$0.42}
& \makecell{31.12\\[-1pt]\footnotesize$\pm$0.52}
& 0.001
& 0.064
& 0.008
\\
\cline{2-7}

& RepTC+IAT
& \makecell{77.89\\[-1pt]\footnotesize$\pm$0.13}
& \makecell{86.55\\[-1pt]\footnotesize$\pm$0.33}
& 0.059
& 2.553
& 0.320
\\
\cline{2-7}

& RepTC+Size+IAT
& \makecell{79.58\\[-1pt]\footnotesize$\pm$1.55}
& \makecell{87.43\\[-1pt]\footnotesize$\pm$0.49}
& 0.060
& 2.553
& 0.321
\\
\cline{2-7}

& \textbf{RepTC}
& \makecell{\textbf{96.21}\\[-1pt]\footnotesize\textbf{$\pm$0.51}}
& \makecell{\textbf{95.33}\\[-1pt]\footnotesize\textbf{$\pm$0.48}}
& 0.055
& 2.553
& 0.306
\\
\hline

\multirow{5}{*}{\makecell{\textbf{USTC-}\\\textbf{TFC2016}}}

& IAT-only
& \makecell{23.11\\[-1pt]\footnotesize$\pm$2.18}
& \makecell{30.14\\[-1pt]\footnotesize$\pm$1.37}
& 0.001
& 0.096
& 0.011
\\
\cline{2-7}

& Size+IAT
& \makecell{59.14\\[-1pt]\footnotesize$\pm$2.40}
& \makecell{66.39\\[-1pt]\footnotesize$\pm$1.17}
& 0.002
& 0.096
& 0.012
\\
\cline{2-7}

& RepTC+IAT
& \makecell{98.74\\[-1pt]\footnotesize$\pm$0.17}
& \makecell{93.21\\[-1pt]\footnotesize$\pm$0.45}
& 0.033
& 1.520
& 0.499
\\
\cline{2-7}

& RepTC+Size+IAT
& \makecell{98.65\\[-1pt]\footnotesize$\pm$0.16}
& \makecell{92.91\\[-1pt]\footnotesize$\pm$0.42}
& 0.033
& 1.520
& 0.499
\\
\cline{2-7}

& \textbf{RepTC}
& \makecell{\textbf{99.62}\\[-1pt]\footnotesize\textbf{$\pm$0.02}}
& \makecell{\textbf{99.42}\\[-1pt]\footnotesize\textbf{$\pm$0.03}}
& 0.029
& 1.520
& 0.483
\\
\hline

\multirow{5}{*}{\makecell{\textbf{Edge-}\\\textbf{IIoTset}}}

& IAT-only
& \makecell{37.24\\[-1pt]\footnotesize$\pm$11.86}
& \makecell{10.17\\[-1pt]\footnotesize$\pm$1.01}
& 0.002
& 0.096
& 0.012
\\
\cline{2-7}

& Size+IAT
& \makecell{67.26\\[-1pt]\footnotesize$\pm$0.08}
& \makecell{51.30\\[-1pt]\footnotesize$\pm$2.73}
& 0.002
& 0.096
& 0.012
\\
\cline{2-7}

& RepTC+IAT
& \makecell{\textbf{99.97}\\[-1pt]\footnotesize\textbf{$\pm$0.00}}
& \makecell{96.22\\[-1pt]\footnotesize$\pm$1.16}
& 0.028
& 1.144
& 0.277
\\
\cline{2-7}

& RepTC+Size+IAT
& \makecell{\textbf{99.97}\\[-1pt]\footnotesize\textbf{$\pm$0.00}}
& \makecell{95.80\\[-1pt]\footnotesize$\pm$1.14}
& 0.028
& 1.144
& 0.278
\\
\cline{2-7}

& \textbf{RepTC}
& \makecell{\textbf{99.97}\\[-1pt]\footnotesize\textbf{$\pm$0.02}}
& \makecell{\textbf{98.29}\\[-1pt]\footnotesize\textbf{$\pm$0.18}}
& 0.025
& 1.144
& 0.262
\\
\hline

\end{tabular}
\end{adjustbox}
\end{table}

\subsubsection{Architecture-Family Transfer}
\label{sec:architecture_transfer}

\added{We further evaluate the representation-aware search with a
non-convolutional architecture space. The convolutional blocks are replaced by
multilayer perceptron (MLP) blocks, while the preprocessing strategies \(S_h\),
input-length range \(L_s\), hardware constraints, training protocol, and
evolutionary search procedure remain unchanged. The MLP architecture space
varies network depth, the number of hidden units per layer, and dropout.}

\added{Table~\ref{tab:architecture_transfer} compares the resulting MLP-based
configuration with the CNN-based RepTC solution. The MLP search selects
\(S_h=S_1\), \(L_s=262\), and a three-layer architecture with 68, 85, and 123
hidden units. It reaches 89.92\% accuracy with a smaller model and
computational footprint than the CNN-based configuration, while the complete
search requires 9.77\,h.}

\begin{table}[htbp]
\centering
\caption{\added{Architecture transfer on ISCX VPN-nonVPN.}}
\label{tab:architecture_transfer}

\renewcommand{\arraystretch}{0.90}
\small
\setlength{\tabcolsep}{3.0pt}

\begin{adjustbox}{max width=\columnwidth}
\begin{tabular}{
>{\centering\arraybackslash}c|
>{\centering\arraybackslash}c|
>{\centering\arraybackslash}c||
>{\centering\arraybackslash}c|
>{\centering\arraybackslash}c||
>{\centering\arraybackslash}c|
>{\centering\arraybackslash}c|
>{\centering\arraybackslash}c||
>{\centering\arraybackslash}c
}
\hline

\textbf{Arch.}
& \(\mathbf{S_h}\)
& \(\mathbf{L_s}\)
& \makecell{\textbf{Acc.}\\(\%)}
& \makecell{\textbf{F1\textsubscript{m}}\\(\%)}
& \makecell{\textbf{Par.}\\(K)}
& \makecell{\textbf{M-Tens.}\\(K)}
& \makecell{\textbf{FLOPs}\\(K)}
& \makecell{\textbf{Wall-clock}\\(h)}
\\
\hline

\textbf{1D-CNN}
& \(S_4\)
& 218
& \makecell{\textbf{96.21}\\[-1pt]\footnotesize\textbf{$\pm$0.51}}
& \makecell{\textbf{95.33}\\[-1pt]\footnotesize\textbf{$\pm$0.48}}
& 54.67
& 2.55
& 305.87
& 18.18
\\
\hline

MLP
& \(S_1\)
& 262
& \makecell{89.92\\[-1pt]\footnotesize$\pm$0.60}
& \makecell{89.12\\[-1pt]\footnotesize$\pm$0.64}
& 36.52
& 0.26
& 72.81
& 9.77
\\
\hline

\end{tabular}
\end{adjustbox}
\end{table}

\added{The CNN-based configuration achieves higher predictive performance,
whereas the MLP search produces a substantially different hardware-feasible
solution with adapted \(S_h\) and \(L_s\).}

\subsection{Robustness Analysis}

\subsubsection{Security-Boundary Stress Test}
\label{sec:w5_robustness}

\added{The final RepTC architectures were evaluated under bounded input
morphing on held-out test data to assess sensitivity to byte-insertion
perturbations while preserving the original traffic labels.
\superadded{The analysis first considers models trained on clean data and then
evaluates perturbation-aware training as a mitigation.} The morphing operation
was implemented through controlled byte insertion at reconstructed packet
boundaries. Two perturbations were considered: zero-byte padding, which
inserts \texttt{0x00} values, and random-byte padding, which inserts uniformly
sampled byte values. The insertion budget was defined as a fraction \(\rho\) of
the useful retained session representation, with
\(\rho\in\{5,10,20,30\}\%\); the resulting sequence was then cropped to the
model-specific input length \(L_s\).}

\added{Table~\ref{tab:w5_robustness} summarizes the robustness results for
the three RepTC models. For each dataset, the clean reference is reported
together with zero-byte and random-byte perturbations at the four insertion
ratios. Accuracy and macro-F1 are reported as mean \(\pm\) standard deviation
over ten runs.}

\added{The stress test reveals a consistent sensitivity to input morphing across
the three datasets. Random-byte padding is generally more disruptive than zero
padding, showing that both the amount and the content of inserted bytes
influence the prediction. ISCX and USTC experience clear degradation under both
perturbation types, while Edge-IIoTset shows the strongest reduction in
macro-F1. On Edge, accuracy remains comparatively high even when macro-F1 drops
sharply, consistent with the strong class imbalance discussed in
Section~\ref{input_sensitivity}; the perturbation therefore affects classes
unevenly rather than producing a uniform loss of performance. The response is
not strictly monotonic at every insertion ratio, but all perturbed conditions
remain below their corresponding clean references, showing that clean-data
training alone provides limited robustness to controlled input morphing and
motivating the mitigation evaluated next.}

\begin{table}[htbp]
\centering
\caption{\added{RepTC robustness under bounded input morphing.}}
\label{tab:w5_robustness}

\renewcommand{\arraystretch}{0.85}
\small
\setlength{\tabcolsep}{3.0pt}

\begin{adjustbox}{max width=\columnwidth}
\begin{tabular}{
>{\centering\arraybackslash}c|
>{\centering\arraybackslash}c||
>{\centering\arraybackslash}c||
>{\centering\arraybackslash}c|
>{\centering\arraybackslash}c|
>{\centering\arraybackslash}c|
>{\centering\arraybackslash}c
}
\hline

\multirow{2}{*}{\textbf{Dataset}}
& \multirow{2}{*}{\textbf{Perturbation}}
& \multirow{2}{*}{\textbf{Clean}}
& \multicolumn{4}{c}{\textbf{Insertion ratio} \(\boldsymbol{\rho}\)}
\\
\cline{4-7}

&
&
& \textbf{5\%}
& \textbf{10\%}
& \textbf{20\%}
& \textbf{30\%}
\\
\hline

\multicolumn{7}{c}{\textit{(a) Accuracy (\%)}} \\
\hline

\multirow{2}{*}{\makecell{\textbf{ISCX}\\\textbf{VPN-nonVPN}}}
& Zero padding
& \multirow{2}{*}{
  \makecell{\textbf{96.21}\\[-1pt]\footnotesize\textbf{$\pm$0.51}}
}
& \makecell{74.36\\[-1pt]\footnotesize$\pm$3.10}
& \makecell{73.61\\[-1pt]\footnotesize$\pm$3.88}
& \makecell{74.80\\[-1pt]\footnotesize$\pm$3.59}
& \makecell{75.02\\[-1pt]\footnotesize$\pm$4.52}
\\
\cline{2-2}\cline{4-7}

& Random padding
&
& \makecell{70.04\\[-1pt]\footnotesize$\pm$4.18}
& \makecell{71.29\\[-1pt]\footnotesize$\pm$3.20}
& \makecell{70.83\\[-1pt]\footnotesize$\pm$3.01}
& \makecell{69.77\\[-1pt]\footnotesize$\pm$2.98}
\\
\hline

\multirow{2}{*}{\makecell{\textbf{USTC-}\\\textbf{TFC2016}}}
& Zero padding
& \multirow{2}{*}{
  \makecell{\textbf{99.62}\\[-1pt]\footnotesize\textbf{$\pm$0.02}}
}
& \makecell{91.98\\[-1pt]\footnotesize$\pm$0.57}
& \makecell{91.87\\[-1pt]\footnotesize$\pm$0.65}
& \makecell{91.55\\[-1pt]\footnotesize$\pm$1.00}
& \makecell{91.82\\[-1pt]\footnotesize$\pm$0.72}
\\
\cline{2-2}\cline{4-7}

& Random padding
&
& \makecell{90.55\\[-1pt]\footnotesize$\pm$0.56}
& \makecell{89.32\\[-1pt]\footnotesize$\pm$0.83}
& \makecell{88.36\\[-1pt]\footnotesize$\pm$0.89}
& \makecell{87.23\\[-1pt]\footnotesize$\pm$0.53}
\\
\hline

\multirow{2}{*}{\makecell{\textbf{Edge-}\\\textbf{IIoTset}}}
& Zero padding
& \multirow{2}{*}{
  \makecell{\textbf{99.97}\\[-1pt]\footnotesize\textbf{$\pm$0.02}}
}
& \makecell{98.91\\[-1pt]\footnotesize$\pm$0.11}
& \makecell{98.92\\[-1pt]\footnotesize$\pm$0.10}
& \makecell{98.93\\[-1pt]\footnotesize$\pm$0.17}
& \makecell{99.17\\[-1pt]\footnotesize$\pm$0.25}
\\
\cline{2-2}\cline{4-7}

& Random padding
&
& \makecell{97.01\\[-1pt]\footnotesize$\pm$1.54}
& \makecell{98.35\\[-1pt]\footnotesize$\pm$0.25}
& \makecell{97.59\\[-1pt]\footnotesize$\pm$0.79}
& \makecell{93.68\\[-1pt]\footnotesize$\pm$2.28}
\\
\hline

\multicolumn{7}{c}{\textit{(b) Macro-F1 (\%)}} \\
\hline

\multirow{2}{*}{\makecell{\textbf{ISCX}\\\textbf{VPN-nonVPN}}}
& Zero padding
& \multirow{2}{*}{
  \makecell{\textbf{95.33}\\[-1pt]\footnotesize\textbf{$\pm$0.48}}
}
& \makecell{80.70\\[-1pt]\footnotesize$\pm$0.72}
& \makecell{80.20\\[-1pt]\footnotesize$\pm$1.49}
& \makecell{81.50\\[-1pt]\footnotesize$\pm$1.56}
& \makecell{80.89\\[-1pt]\footnotesize$\pm$1.66}
\\
\cline{2-2}\cline{4-7}

& Random padding
&
& \makecell{75.84\\[-1pt]\footnotesize$\pm$2.94}
& \makecell{77.35\\[-1pt]\footnotesize$\pm$1.83}
& \makecell{77.67\\[-1pt]\footnotesize$\pm$2.06}
& \makecell{76.72\\[-1pt]\footnotesize$\pm$1.05}
\\
\hline

\multirow{2}{*}{\makecell{\textbf{USTC-}\\\textbf{TFC2016}}}
& Zero padding
& \multirow{2}{*}{
  \makecell{\textbf{99.42}\\[-1pt]\footnotesize\textbf{$\pm$0.03}}
}
& \makecell{81.59\\[-1pt]\footnotesize$\pm$0.62}
& \makecell{80.88\\[-1pt]\footnotesize$\pm$0.41}
& \makecell{79.58\\[-1pt]\footnotesize$\pm$1.54}
& \makecell{80.17\\[-1pt]\footnotesize$\pm$1.52}
\\
\cline{2-2}\cline{4-7}

& Random padding
&
& \makecell{78.73\\[-1pt]\footnotesize$\pm$0.74}
& \makecell{76.41\\[-1pt]\footnotesize$\pm$0.66}
& \makecell{73.99\\[-1pt]\footnotesize$\pm$0.67}
& \makecell{73.12\\[-1pt]\footnotesize$\pm$0.44}
\\
\hline

\multirow{2}{*}{\makecell{\textbf{Edge-}\\\textbf{IIoTset}}}
& Zero padding
& \multirow{2}{*}{
  \makecell{\textbf{98.29}\\[-1pt]\footnotesize\textbf{$\pm$0.18}}
}
& \makecell{60.17\\[-1pt]\footnotesize$\pm$2.51}
& \makecell{60.23\\[-1pt]\footnotesize$\pm$2.57}
& \makecell{60.42\\[-1pt]\footnotesize$\pm$4.83}
& \makecell{63.85\\[-1pt]\footnotesize$\pm$6.94}
\\
\cline{2-2}\cline{4-7}

& Random padding
&
& \makecell{54.87\\[-1pt]\footnotesize$\pm$2.17}
& \makecell{49.75\\[-1pt]\footnotesize$\pm$2.77}
& \makecell{44.77\\[-1pt]\footnotesize$\pm$6.02}
& \makecell{40.83\\[-1pt]\footnotesize$\pm$6.03}
\\
\hline

\end{tabular}
\end{adjustbox}
\end{table}

\superadded{We further examine whether this sensitivity can be mitigated through
perturbation-aware training while keeping the selected RepTC architectures and
representations unchanged. Training combines 50\% clean samples with 25\%
zero-byte and 25\% random-byte insertions, with \(\rho\) sampled from the same
tested ratios; validation remains clean. The models are then evaluated under
the same clean and perturbed test conditions, and
Table~\ref{tab:w5_mitigation} reports the resulting macro-F1.}

\superadded{
Perturbation-aware training substantially improves robustness under
both perturbation types. The effect is strongest under random-byte padding,
where the clean-trained models degrade most. There, at \(\rho=30\%\), macro-F1
improves by 12.26, 19.97, and 41.91 percentage points on ISCX, USTC, and
Edge-IIoTset, respectively, while the perturbation-aware models remain more
stable across the tested insertion ratios. The improvement comes with lower
clean macro-F1, indicating a trade-off between clean performance and
robustness. The results show that this sensitivity can be
mitigated without changing the selected RepTC architectures or
representations.}

\begin{table}[htbp]
\centering
\caption{\superadded{Macro-F1 under perturbation-aware training.}}
\label{tab:w5_mitigation}

\renewcommand{\arraystretch}{0.85}
\small
\setlength{\tabcolsep}{3.0pt}

\begin{adjustbox}{max width=\columnwidth}
\begin{tabular}{
>{\centering\arraybackslash}c|
>{\centering\arraybackslash}c||
>{\centering\arraybackslash}c||
>{\centering\arraybackslash}c|
>{\centering\arraybackslash}c|
>{\centering\arraybackslash}c|
>{\centering\arraybackslash}c
}
\hline

\multirow{2}{*}{\textbf{Dataset}}
& \multirow{2}{*}{\textbf{Perturbation}}
& \multirow{2}{*}{\textbf{Clean}}
& \multicolumn{4}{c}{\textbf{Insertion ratio} \(\boldsymbol{\rho}\)}
\\
\cline{4-7}

&
&
& \textbf{5\%}
& \textbf{10\%}
& \textbf{20\%}
& \textbf{30\%}
\\
\hline

\multirow{2}{*}{\makecell{\textbf{ISCX}\\\textbf{VPN-nonVPN}}}
& Zero padding
& \multirow{2}{*}{\makecell{\textbf{91.43}\\[-1pt]\footnotesize$\pm$0.41}}
& \makecell{89.49\\[-1pt]\footnotesize$\pm$0.37}
& \makecell{89.69\\[-1pt]\footnotesize$\pm$0.44}
& \makecell{89.78\\[-1pt]\footnotesize$\pm$0.52}
& \makecell{89.57\\[-1pt]\footnotesize$\pm$0.19}
\\
\cline{2-2}\cline{4-7}
& Random padding
&
& \makecell{89.43\\[-1pt]\footnotesize$\pm$0.32}
& \makecell{89.38\\[-1pt]\footnotesize$\pm$0.28}
& \makecell{89.18\\[-1pt]\footnotesize$\pm$0.23}
& \makecell{88.98\\[-1pt]\footnotesize$\pm$0.13}
\\
\hline

\multirow{2}{*}{\makecell{\textbf{USTC-}\\\textbf{TFC2016}}}
& Zero padding
& \multirow{2}{*}{\makecell{\textbf{93.57}\\[-1pt]\footnotesize$\pm$0.04}}
& \makecell{93.23\\[-1pt]\footnotesize$\pm$0.07}
& \makecell{93.24\\[-1pt]\footnotesize$\pm$0.14}
& \makecell{93.18\\[-1pt]\footnotesize$\pm$0.12}
& \makecell{93.15\\[-1pt]\footnotesize$\pm$0.13}
\\
\cline{2-2}\cline{4-7}
& Random padding
&
& \makecell{93.22\\[-1pt]\footnotesize$\pm$0.15}
& \makecell{93.20\\[-1pt]\footnotesize$\pm$0.12}
& \makecell{93.11\\[-1pt]\footnotesize$\pm$0.07}
& \makecell{93.09\\[-1pt]\footnotesize$\pm$0.12}
\\
\hline

\multirow{2}{*}{\makecell{\textbf{Edge-}\\\textbf{IIoTset}}}
& Zero padding
& \multirow{2}{*}{\makecell{\textbf{85.79}\\[-1pt]\footnotesize$\pm$1.01}}
& \makecell{85.71\\[-1pt]\footnotesize$\pm$0.93}
& \makecell{85.69\\[-1pt]\footnotesize$\pm$1.17}
& \makecell{85.65\\[-1pt]\footnotesize$\pm$1.02}
& \makecell{85.73\\[-1pt]\footnotesize$\pm$0.95}
\\
\cline{2-2}\cline{4-7}
& Random padding
&
& \makecell{84.62\\[-1pt]\footnotesize$\pm$1.60}
& \makecell{83.21\\[-1pt]\footnotesize$\pm$1.31}
& \makecell{83.22\\[-1pt]\footnotesize$\pm$1.57}
& \makecell{82.74\\[-1pt]\footnotesize$\pm$0.74}
\\
\hline

\end{tabular}
\end{adjustbox}
\end{table}

\subsubsection{Temporal Distribution-Shift Analysis}
\label{sec:temporal_robustness}

\added{\superadded{Temporal non-stationarity is an important challenge in ETC,
as traffic patterns evolve over time and models trained on earlier captures
may generalize less effectively to later traffic~\cite{malekghaini2023deep}.}
We therefore evaluate the selected RepTC model under a chronological holdout
to assess generalization under temporal distribution shift.
Among the three datasets, ISCX VPN-nonVPN provides a basis for chronological
evaluation, as 10 of its 11 classes span multiple PCAPs and several classes
cover extended capture periods. In contrast, Edge-IIoTset contains a single
PCAP per class, while most USTC-TFC2016 classes are represented by a single
PCAP and temporal coverage is highly uneven across classes; chronological
splits on these datasets would therefore provide a weaker measure of temporal
shift.}

\added{Here, sessions within each class are ordered by capture
time and split chronologically into the earliest 70\% for training, the next
10\% for validation, and the latest 20\% for testing. The final RepTC
architecture and representation are kept unchanged, and the model is retrained
using only earlier training samples. Over ten runs, the
selected configuration achieves \(64.42 \pm 2.54\%\) accuracy,
\(73.22 \pm 1.92\%\) macro-F1, and \(62.39 \pm 1.91\%\) weighted-F1 on the
later test samples.}


\superadded{The reduction relative to the main evaluation indicates reduced generalization to temporally later traffic within the available ISCX capture period and motivates temporal robustness as a direction for future
optimization.}

\section{Deployment and Real-Time Evaluation} \label{deployment}

\subsection{Setup and Methodology}

Tests on real-time feasibility covered two deployment scenarios: (i) full pipeline execution on a Raspberry Pi 3 Model B+, and (ii) a hybrid setup where the Pi acted as a live traffic gateway, forwarding processed sessions to \deleted{an STM32 Nucleo-F401RE}\superadded{a resource-constrained MCU} for inference. \superadded{The hybrid scenario was evaluated on both the STM32 Nucleo-F401RE (Cortex-M4) and the XIAO ESP32-C3 (RISC-V), providing heterogeneous MCU deployment targets.} These setups reflected specific tradeoff situations: the former setting sustained ultra-low-latency inference \deleted{on one embedded platform}\superadded{on the edge gateway}, whereas the latter one involved offloading to a resource-constrained microcontroller. Table~\ref{tab:hw_specs} summarizes the hardware specs of \deleted{both}\superadded{the three} platforms, in terms of computational resources (Central Processing Unit (CPU), RAM) and interfaces.

\begin{table}[b]
    \centering
    \caption{Deployment platform specifications.}
    \label{tab:hw_specs}
    \renewcommand{\arraystretch}{1.05}
    \scriptsize

    \resizebox{\columnwidth}{!}{%
    \begin{tabular}{c|c|c|c}
        \hline
        \textbf{Specification} &
        \makecell{\textbf{Raspberry Pi}\\\textbf{3B+}} &
        \makecell{\textbf{STM32 Nucleo-}\\\textbf{F401RE}} &
        \superadded{\makecell{\textbf{XIAO}\\\textbf{ESP32-C3}}}
        \\
        \hline
        CPU/Core & Cortex-A53 & Cortex-M4 & \superadded{RISC-V} \\
        \hline
        Clock & 4$\times$1.4\,GHz & 84\,MHz & \superadded{160\,MHz} \\
        \hline
        RAM & 1\,GB LPDDR2 & 96\,KB SRAM & \superadded{400\,KB SRAM} \\
        \hline
        Flash / Storage & microSD (ext.) & 512\,KB Flash & \superadded{4\,MB Flash} \\
        \hline
        OS / SDK & Raspbian Lite & STM32Cube.AI & \superadded{ESP-IDF / TFLM} \\
        \hline
        Connectivity & Ethernet, UART & UART & \superadded{USB/UART} \\
        \hline
        \makecell{Measured\\voltage} & 4.75\,V & 4.55\,V & \superadded{5.07\,V} \\
        \hline
    \end{tabular}%
    }
\end{table}


The RepTC models were quantized to 8-bit integer precision using quantization-aware training (QAT), then exported as INT8 TensorFlow Lite models. \deleted{They were converted to C code via STM32Cube.AI for STM32 deployment.}\superadded{For deployment, STM32Cube.AI generated the STM32 implementation, while the ESP32 implementation used TensorFlow Lite for Microcontrollers (TFLM) under ESP-IDF.} QAT was adopted to fine-tune each model under simulated INT8 constraints before deployment, thereby mitigating quantization-induced accuracy loss. For each Float32 model from the ten independent runs in Section~\ref{results}, a corresponding QAT-based INT8 model was generated to assess repeatability and consistency.
 

The gateway was then used to capture real-time traffic using Scapy, yielding bidirectional sessions in 5-dimensional tuples (source/destination IPs and ports, protocol). Each session was processed once it reached the target input length \(L_s\), or flushed after 10\,s of inactivity. The strategy \(S_h\) preprocessed the finalized byte stream, which was converted to an 8-bit format and passed either to the local inference model or \deleted{offloaded to the STM32}\superadded{offloaded to an MCU node (STM32 or ESP32)} via serial communication to obtain the predicted label. The gateway was configured to maintain an in-memory buffer of up to 1000 concurrent sessions; if this limit was reached, the least-recently-active session was evicted and underwent standard processing.


A USB power meter (3--9\,V, 0--3\,A, \(\pm1\%\) accuracy) was placed between each device and its power source to record current and voltage. Power was computed as \(P=V\times I\), and energy per session was estimated by multiplying average power 
by the duration of the corresponding measured
operation. The reported values refer to the additional consumption above the idle baseline of each platform; on the Raspberry Pi, the baseline was measured after system stabilization with background services minimized and no extra user processes running. System-level power or energy can be obtained by adding the idle draw to these values.

\subsection{Deployment Results and Analysis}

\begin{figure}[t]
    \centering
    \includegraphics[width=\columnwidth]{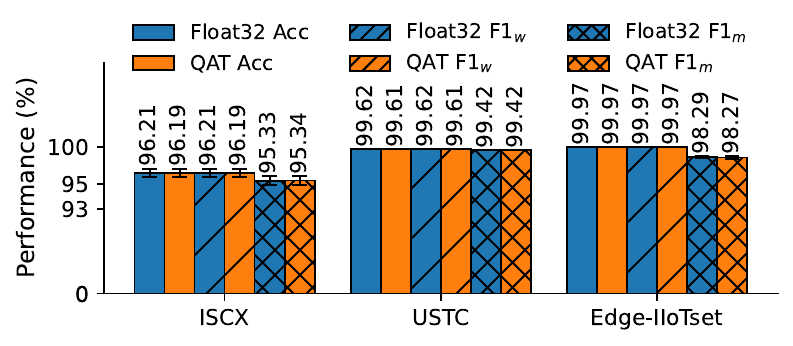}
        \caption{Quantization performance of RepTC models.}
    \label{fig:quantization_accuracy}
\end{figure}

Figure~\ref{fig:quantization_accuracy} compares the accuracy (Acc), weighted F1-score (F1\textsubscript{w}), and macro F1-score (F1\textsubscript{m}) of the Float32 and INT8 variants on the ISCX, USTC, and Edge-IIoTset datasets. Each bar group corresponds to a dataset-specific model and shows accuracy (solid) and F1-scores (hatched) before and after quantization, together with standard deviations over ten runs.

QAT caused negligible degradation across all three datasets. The Float32 and INT8 results were nearly identical on USTC and Edge-IIoTset, while ISCX also remained stable, with only marginal differences in accuracy and F1-scores.
These results confirm that RepTC designs remain stable under INT8 quantization and are well suited to embedded deployment.

\begin{figure}[h]
    \centering
    \includegraphics[width=\columnwidth]{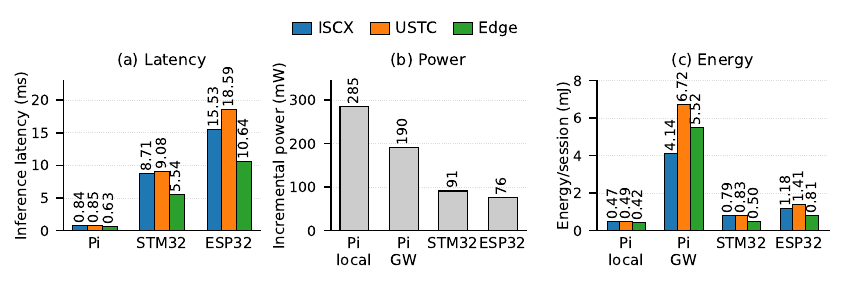}
    \caption{\superadded{Deployment metrics across models and scenarios.}}
    \label{fig:deployment_metrics}
\end{figure}

Following this, Figure~\ref{fig:deployment_metrics} \deleted{gives deployment metrics for the various scenarios, spanning inference latency, power, and energy for both the Pi and STM32. Solid bars represent Pi-only execution; hatched bars indicate Pi+STM32 schemes, in which the STM32 performs inference. Inference latency relates to Pi execution in the Pi-only setup, and STM32 execution in the Pi+STM32 setup. Power and energy are measured during preprocessing and inference (Pi-only), or preprocessing and UART wait (Pi+STM32). STM32 metrics only consider inference.}\superadded{summarizes the deployment measurements across the evaluated scenarios. Figure~\ref{fig:deployment_metrics}(a) reports inference latency on the Pi, STM32, and ESP32; Fig.~\ref{fig:deployment_metrics}(b) compares incremental power, which remained stable across the three dataset-specific models and is therefore shown once per device role; and Fig.~\ref{fig:deployment_metrics}(c) reports per-session energy. For the Pi, power and energy correspond to preprocessing and local inference in the Pi-only setup (Pi local), or preprocessing and UART transfer/wait in the offloading setup (Pi GW). STM32 and ESP32 values correspond to MCU-side inference only.}

All values represent averages over 1000 captured sessions and exclude the idle baseline, which was \(\approx0.53\,\mathrm{A}\) at 4.75\,V (2.52\,W) for the Raspberry Pi 3B+, \(\approx0.02\,\mathrm{A}\) at 4.55\,V (91\,mW) for the STM32 Nucleo-F401RE, \superadded{and \(\approx0.018\,\mathrm{A}\) at 5.07\,V (91.3\,mW) for the XIAO ESP32-C3.}

In Pi-only mode, average inference latency was 0.84\,ms (ISCX), 0.85\,ms (USTC), and 0.63\,ms (Edge), with \(S_h\) preprocessing taking \(\sim\)0.8--0.9\,ms per session. Pi power kept stable at 285\,mW, yielding per-session energy consumption of 0.47\,mJ, 0.49\,mJ, and 0.42\,mJ, respectively. 
Profiling also showed low resource usage, between 1.47--2.39\% CPU and 54.5--54.9\,MB RAM, even with the Pi handling full execution.


In the offloading setup, STM32 inference latency was 8.71\,ms for ISCX, 9.08\,ms for USTC, and 5.54\,ms for Edge. \superadded{ESP32 inference ranged from 10.64\,ms to 18.59\,ms across the three models.} \deleted{UART transmission added 20.97\,ms, 34.48\,ms, and 28.20\,ms for session sizes of \(L_s=218\), 530, and 314 bytes, respectively, transferred at a baudrate of 1.5\,Mbps.}\superadded{Both MCU nodes used UART at 1.5\,Mbps, with measured transfer delays of approximately 20.97\,ms, 34.48\,ms, and 28.20\,ms for session sizes of \(L_s=218\), 530, and 314 bytes, respectively.} These values reflect end-to-end transfer with software and handshake overheads, not only the line rate. 

While Pi power dropped to 190\,mW, its energy usage increased due to UART-induced wait time, reaching 4.14\,mJ for ISCX, 6.72\,mJ for USTC, and 5.52\,mJ for Edge. The STM32 power during inference was approximately 91\,mW, with energy costs of 0.79\,mJ, 0.83\,mJ, and 0.50\,mJ, respectively. \superadded{ESP32 inference required 76.05\,mW and 0.81--1.41\,mJ across the three models.} In this gateway-only role, the Pi used 1.59--2.52\% CPU and \(\sim\)49.8\,MB RAM.


Overall, Pi-only mode achieved sub-millisecond inference with minimal energy, making it well-suited for low-latency gateway execution. \deleted{STM32 offloading incurred moderate UART delays, but featured lower peak power and enabled inference on ultra-low-power microcontrollers.}\superadded{MCU offloading incurred UART delays but enabled inference on both Cortex-M4 and RISC-V targets; STM32 provided lower inference latency, while ESP32 operated at lower inference power.} FENXI~\cite{gallo2021fenxi} is, to the best of our knowledge, among the few approaches reporting real-time TC, with 10--30\,ms latency on a 30\,W Peripheral Component Interconnect Express (PCIe) TPU at 100\,Gbps. This, however, relies on costly server-class accelerators (tens of Watts), suited to datacenter workloads rather than embedded devices. In contrast, prior work~\cite{our_chehade2025iscc,our_chehade2025tnsm} reported MCU-side inference with considerably higher latency (31--115\,ms) and energy consumption (7.9--29.1\,mJ), i.e., several times higher than those achieved by the proposed approach.

\section{Conclusion}

The RepTC approach jointly optimizes model structure, session length, and header preprocessing to design efficient traffic classifiers under IoT constraints.
Results show that these three dimensions significantly affect both performance and resource consumption.
The resulting models are compact, accurate, and deployable on real edge devices. On encrypted and industrial IoT datasets, the method achieves up to 99.97\% accuracy. In deployment tests,  full prototypes supported real-time session processing and inference, running reliably on low-power platforms such as  Raspberry Pi and microcontrollers. 
\superadded{Robustness analyses also reveal sensitivity to temporal distribution
shift and input morphing, while perturbation-aware training markedly improves
resilience to the latter. Overall,} the findings highlight that input representation should be treated as a key design variable, alongside architecture, in hardware-constrained TC.
\deleted{For future work, we plan to explore adversarial robustness,
explainability for constrained models, and to assess how representation-aware
optimization generalizes to evolving traffic distributions and cross-environment
settings; we also aim to extend this approach to federated learning across
distributed nodes, thus advancing TC toward secure, privacy-aware IoT
analytics.}\superadded{For future work, we plan to investigate robustness-aware RepTC
search that accounts for input perturbations and temporal or cross-environment
shifts during optimization; we further plan to evaluate RepTC on dedicated
longitudinal traffic data. We also aim to study explainability for constrained
models and federated learning across distributed nodes, toward secure,
privacy-aware IoT analytics.}


\bibliography{references}



\end{document}